\pdfoutput=1
\documentclass[%
 reprint,
nofootinbib,
amsmath,amssymb,
aps,
]{revtex4-2}

\usepackage{graphicx}
\usepackage{dcolumn}
\usepackage{bm}

\usepackage{amsmath,amssymb,amsbsy,amstext,amsthm,simplewick,amsfonts,graphicx}
\usepackage{mathrsfs}
\usepackage{graphicx}
\usepackage{wrapfig}
\usepackage{upgreek}
\usepackage{bm} 
\usepackage{framed}
\usepackage{bbm}
\usepackage{textcomp}
\usepackage{adjustbox}
\usepackage{makecell}
\usepackage{tcolorbox}
\usepackage{empheq}
\usepackage[normalem]{ulem}
\usepackage{enumitem}
\usepackage{braket} 
\usepackage{array}
\usepackage{dsfont}
\usepackage{physics}
\usepackage{ulem}
\usepackage{xcolor}
\usepackage{caption}
\usepackage{subcaption}
\usepackage{tocloft}
\usepackage{tikz}
\usepackage{xcolor}
\usetikzlibrary{patterns}

\usepackage{hyperref}
\hypersetup{colorlinks=true,
	breaklinks=true,
	pdfstartview=Fit,
	linkcolor=blue2,
	citecolor=red2,
	urlcolor=blue
}

\definecolor{red2}{RGB}{214, 39, 40}
\definecolor{green2}{RGB}{0,170,0}
\definecolor{blue2}{RGB}{0,100,200}
\definecolor{magenta2}{RGB}{191,64,191}
\definecolor{purple2}{RGB}{112,48,160}
\definecolor{orange2}{RGB}{255,192,0}

\newcommand{\textOrange}[1]{\textcolor[RGB]{245,166,35}{#1}}
\newcommand{\textGreen}[1]{\textcolor[RGB]{126,211,33}{#1}}
\newcommand{\ex}[1]{\left\langle #1 \right\rangle}

\graphicspath{{Fig/}}

\def\Det{{\rm Det}}

\def\d{\mathrm{d}}

\def\bfk{\mathbf{k}}

\def\bfs{\mathbf{s}}
\def\bft{\mathbf{t}}
\def\bfu{\mathbf{u}}

\begin{document}

\preprint{APS/123-QED}

\title{A Correlator-Wavefunction Duality for Primordial Perturbations \\ and the  factorisation among correlators}

\title{From Wavefunction Coefficients to Cosmological Correlators, and Back Again:\\ \vspace{0.2cm} {\textit{duality relation and the  factorisation among correlators}}}

\title{There and Back Again: Mapping and Factorising Cosmological Observables}

\author{David Stefanyszyn}
\email{david.stefanyszyn@nottingham.ac.uk}
\affiliation{
School of Mathematical Sciences \& School of Physics and Astronomy, University of Nottingham,\\ University Park, Nottingham, NG7 2RD, UK}

\author{Xi Tong}%
\email{xt246@cam.ac.uk}
\affiliation{%
Department of Applied Mathematics and Theoretical Physics, University of Cambridge,\\Wilberforce Road, Cambridge, CB3 0WA, UK
}%

\author{Yuhang Zhu}
\email{yhzhu@ibs.re.kr}
\affiliation{
Cosmology, Gravity and Astroparticle Physics Group, Center for Theoretical Physics of the Universe,\\
        Institute for Basic Science, Daejeon, 34126, Korea
}%

\date{\today}

\begin{abstract}
Cosmological correlators encode invaluable information about the wavefunction of the primordial universe. In this \textit{letter} we present a duality between correlators and wavefunction coefficients that is valid to all orders in the loop expansion and manifests itself as a $\mathbb{Z}_4$ symmetry. To demonstrate the power of the duality, we derive a correlator-to-correlator factorisation (CCF) formula for the parity-odd part of cosmological correlators that relates $n$-point observables to lower-point ones via a series of diagrammatic cuts. These relations serve as the first example of physically testable cutting rules as they involve observables defined for arbitrary physical kinematics. We further show how the duality allows us to translate the cosmological optical theorem for wavefunction coefficients into statements about cosmological correlators.
\end{abstract}

\maketitle


\section{Introduction}
One of the ultimate goals of physics is to understand the laws of nature at the beginning of time. A cosmologist's approach to this problem is to measure spatial correlations in the Cosmic Microwave Background radiation and the Large Scale Structure of the universe. These correlations are seeded by primordial \textit{cosmological correlators} of quantum fields evolving during inflation, with their momentum dependence encoding the secrets of the underlying physics at play during the universe's first moments. For example, the soft limit of correlators probes the cosmic expansion history \cite{Chen:2015lza,Wang:2020aqc} and the inflationary particle spectrum \cite{Chen:2009zp,Baumann:2011nk,Noumi:2012vr,Arkani-Hamed:2015bza,Lee:2016vti,Jazayeri:2022kjy,Pimentel:2022fsc,Cabass:2024wob,Chakraborty:2023qbp,Chen:2022vzh,Tong:2022cdz,Reece:2022soh,Wang:2020ioa,Kumar:2019ebj,Sohn:2024xzd,McCulloch:2024hiz,Craig:2024qgy,Chakraborty:2023qbp,Jazayeri:2023xcj,Qin:2022lva,Chen:2022vzh,Bodas:2020yho,Wang:2019gbi,Meerburg:2016zdz,Chen:2016uwp,Bordin:2018pca,Xianyu:2023ytd,Aoki:2024uyi,Chakraborty:2023eoq,Chen:2023txq}, the equilateral limit probes higher-dimensional self-interactions of the inflaton \cite{Cheung:2007st,Chen:2006nt,Baumann:2011su,Creminelli:2003iq,Seery:2005wm}, while the collinear limit probes the initial state \cite{Chen:2006nt,Holman:2007na,Meerburg:2009ys,Ganc:2011dy,Flauger:2013hra,Green:2020whw} and environmental effects \cite{Salcedo:2024smn}. Understanding the structure of cosmological correlators is therefore of upmost importance in our quest to understand the early universe and therefore physics at extreme energy scales.

In recent years, however, much attention has been paid to more primitive objects, namely \textit{wavefunction coefficients} that encode the inflationary dynamics in the perturbative expansion of the wavefunction of the universe (given that we work on a fixed background geometry, ``field-theoretic wavefunction" might be a more appropriate name) \cite{Maldacena:2002vr,Anninos:2014lwa}. Although these objects are not directly observable, cosmological correlators can be extracted from them by applying the Born rule, and their somewhat simpler kinematic dependence means that constraints from cherished physical principles such as symmetries, locality and unitarity turn out to be more transparent \cite{Arkani-Hamed:2018kmz,Baumann:2019oyu,Baumann:2020dch,Baumann:2021fxj,Albayrak:2023hie,Arkani-Hamed:2017fdk,Benincasa:2022omn, Goodhew:2020hob,Melville:2021lst,Goodhew:2021oqg,Pajer:2020wxk,Jazayeri:2021fvk,Melville:2021lst, Salcedo:2022aal, Meltzer:2021zin,Baumann:2021fxj,Donath:2024utn,Lee:2023kno,Agui-Salcedo:2023wlq,Cespedes:2020xqq,Baumann:2022jpr,DiPietro:2021sjt,Hogervorst:2021uvp}. They also play an important role in defining cosmological amplitudes \cite{Melville:2023kgd,Melville:2024ove,Fan:2024iek,Gomez:2021ujt,Donath:2024utn}, can be used to understand the origin of IR-divergences in de Sitter space \cite{Cespedes:2023aal,Bzowski:2023nef,Gorbenko:2019rza}, contain (boost-breaking) flat-space amplitudes in a certain singular kinematic limit \cite{Maldacena:2011nz,Raju:2012zr,Pajer:2020wnj}, and have neat connections to geometry \cite{Benincasa:2024lxe,Benincasa:2024leu,Arkani-Hamed:2023kig,Arkani-Hamed:2017fdk}.

In this \textit{letter}, we derive a duality between cosmological correlators $B_n$ and the ``physical" part of wavefunction coefficients $\rho_n\equiv \psi_n+\psi_n^\sharp$, where $\sharp$ stands for complex conjugation and momentum reversal \footnote{It is this combination that appears in the square modulus of the wavefunction of the universe and therefore contributes to the Born rule. In momentum space, it reads $\rho_n(\{\mathbf{k}\})= \psi_n(\{\mathbf{k}\})+\psi_n^*(\{-\mathbf{k}\})$ where $\{ \bfk \}$ collectively denotes the external momenta.}, that is valid to all orders in perturbation theory (i.e. to all orders in the loop expansion). We show that in the dictionary that translates the $\{\rho_n\}$ to the $\{B_n\}$, there exists a $\mathbb{Z}_4$ symmetry that \textit{syntactically} swaps $\rho_n\leftrightarrow B_n$ and maps any valid equation to another valid equation within the dictionary. The duality therefore takes us from wavefunction coefficients to cosmological correlators, \textit{and back again}.

We use this duality, in combination with the results of \cite{Stefanyszyn:2023qov} where we showed that parity-odd correlators are factorised, to prove that under a set of mild assumptions, parity-odd correlators of inflatons and gravitons factorise into a \textit{structured sum of products of lower-point correlators}. This factorisation holds for physical kinematics (no need for analytic continuation) and for generic momentum dependence (no need to take specific kinematic limits) and is in principle a relation that can tested with observations. We show that it also has a neat interpretation in terms of diagrammatic cuts. The first non-trivial example of this correlator-to-correlator factorisation (CCF) relates the trispectrum of primordial perturbations to the bispectra involving two curvature perturbations and one additional state with integer spin and a complementary series mass, and the power spectrum of this state. The power of this relation lies in the fact that it maps an observable to a combination of other observables.

We further show the usefulness of our duality in the context of unitarity and the cosmological optical theorem (COT) \cite{Goodhew:2020hob}. The COT is most naturally derived for wavefunction coefficients since unitary time evolution imposes a set of conditions on the wavefunction of the universe \cite{Goodhew:2020hob,Melville:2021lst,Goodhew:2021oqg,Cespedes:2020xqq,Baumann:2021fxj}. It manifests itself as a relation between analyitcally continued wavefunction coefficients. It is desirable, however, to derive conditions on cosmological correlators since these are ultimately the fundamental observables and while some progress has been made in this direction \cite{Donath:2024utn} (see also \cite{Tong:2021wai,Ema:2024hkj,Qin:2023bjk,Qin:2023nhv} for cutting rules that focus on the cosmological collider signal), the full set of conditions has not been derived (even at tree-level). In this paper we show how our duality can play an important role in this regard by converting the COT for wavefunction coefficients into statements for cosmological correlators.

{\bf Notations and conventions}. For conciseness, we use the DeWitt notation \cite{DeWitt:1967ub}, where both field indices and spatial coordinates are abbreviated as a single Latin index as $\varphi^A(\mathbf{x})\equiv \varphi_i$. Contractions are interpreted as $\varphi_i\chi^i\equiv \sum_A\int \d^3x\, \varphi_A(\mathbf{x})\chi^A(\mathbf{x})$. We adopt the following diagrammatic notations:
\begin{align}
     B_n~\equiv~\raisebox{-10pt}{\includegraphics[scale=0.8]{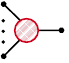}}~,\qquad\rho_n~\equiv~\raisebox{-10pt}{\includegraphics[scale=0.8]{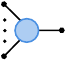}}~.
\end{align}
These blobs represent the abstract notion of correlators and wavefunction coefficients respectively, without specific perturbation theory structures inside. They should be distinguished from the typical Schwinger-Keldysh/Witten diagrams that compute $B_n$ and $\rho_n$. For conciseness, all momentum and tensor indices have been suppressed.

\section{Correlator-wavefunction duality}

We start with a lightning review of the wavefunction approach to primordial perturbations. Consider a set of weakly interacting quantum fields collectively denoted by $\Phi^i(\eta)$ evolving in a classical inflationary spacetime $g_{\mu\nu}=a^2(\eta)\eta_{\mu\nu}$, $a(\eta)=-1/(H\eta)$. In the Schrödinger picture, we define the wavefunction of the universe by projecting the time-evolved Bunch-Davies vacuum onto a field-value eigenstate:
\begin{align}
	\Psi[\varphi]=\langle \varphi|U(\eta_0,-\infty)|\text{BD}\rangle =\int_{\text{BD}}^{\Phi(\eta_0)=\varphi} \mathcal{D}\Phi\, e^{i S[\Phi]} \,.
\end{align}
In practice, the Bunch-Davies vacuum choice means that we integrate over fields that vanish in the far past (with a suitable contour deformation). In perturbation theory, the wavefunction is conventionally parametrised by
\begin{align} \label{WavefunctionExpansion}
\Psi[\varphi]=\exp\left[+\sum_{n=1}^{\infty}\frac{1}{n!}(\psi_n)^{i_1\cdots i_n}\varphi_{i_1}\cdots\varphi_{i_n}\right]\,,
\end{align}
with the coefficients $\psi_n$ computed via Witten diagrammatics. Given our notation, there is an implicit sum over different fields in \eqref{WavefunctionExpansion} thereby ensuring that each wavefunction coefficient has the appropriate normalisation e.g. for $\psi_{3}$ with only two identical fields, we have an overall factor of $\frac{3}{3!} = \frac{1}{2!}$. We unconventionally start the summation with $n=1$. This tadpole term starts out at loop-level and is necessary for the cancellation of the monopole moment in observables.

The full $n$-point correlators of quantum fields are computed by the Born rule as \footnote{Note that in perturbation theory, the path integral is always computed by expanding and truncating the non-Gaussian part of $|\Psi[\varphi]|^2$, with convergence guaranteed by the Gaussian part with $\rho_2<0$.}
\begin{align} \label{BornRule}
    \ex{\varphi_{i_1}\cdots\varphi_{i_n}}=\dfrac{\int \mathcal{D}\varphi\, |\Psi[\varphi]|^2\varphi_{i_1}\cdots\varphi_{i_n}}{\int \mathcal{D}\varphi\, |\Psi[\varphi]|^2} \,.
\end{align}
However, the quantities of more observational relevance are their connected part, which we will denote as $B_n$, in which only one momentum-conserving $\delta$-function is present in momentum space. These are computed from the generating functional
\begin{align}
	Z[J]=\int\mathcal{D}\varphi\, |\Psi[\varphi]|^2 e^{i J^i \varphi_i}\,,\label{generatingFunctional}
\end{align}
by taking derivatives with respect to the auxiliary current i.e.
\begin{align}
	(B_n)_{i_1\cdots i_n}=\left.\frac{\partial}{i \,\partial J^{i_1}}\cdots\frac{\partial}{i \,\partial J^{i_n}}\ln Z[J]\,\right|_{J=0} \,.\label{BnDefDiffForm}
\end{align}
This yields the correct normalisation since the denominator in \eqref{BornRule} is simply $Z[0]$, and we emphasise that the derivatives act on $\ln Z[J]$ rather than $Z[J]$ thereby ensuring we only extract connected correlators. Alternatively, we can integrate \eqref{BnDefDiffForm} to obtain
\begin{align}
	\nonumber&\exp\left[{+\sum_{n=1}^{\infty}\frac{i^n}{n!}(B_n)_{i_1\cdots i_n}J^{i_1}\cdots J^{i_n}}\right]\\
	&=\int\mathcal{D}\varphi\, \exp\left[+\sum_{n=1}^{\infty}\frac{1}{n!}(\rho_n)^{i_1\cdots i_n}\varphi_{i_1}\cdots\varphi_{i_n}\right]e^{i J^i \varphi_i}~,\label{DictIntForm}
\end{align}
where $\rho_n\equiv \psi_n+\psi_n^\sharp$ is the physical part of a wavefunction coefficient. Therefore, the \textit{connected correlators} can be viewed as a Fourier transformation of the physical wavefunction coefficients. Now notice that the two sides of \eqref{DictIntForm} take completely analogous forms. We can therefore perform an inverse Fourier transformation to rewrite \eqref{DictIntForm} as
\begin{align}
\nonumber&\exp\left[+\sum_{n=1}^{\infty}\frac{1}{n!}(\rho_n)^{i_1\cdots i_n}\phi_{i_1}\cdots\phi_{i_n}\right]\\
	&=\int\mathcal{D}J\, \exp\left[{+\sum_{n=1}^{\infty}\frac{i^n}{n!}(B_n)_{i_1\cdots i_n}J^{i_1}\cdots J^{i_n}}\right]e^{-i J^i \phi_i} \,.\label{DictIntForm2}
\end{align}
After a change of the dummy variable $J\to -J$, we obtain
\begin{align}
\nonumber&\exp\left[+\sum_{n=1}^{\infty}\frac{1}{n!}(\rho_n)^{i_1\cdots i_n}\phi_{i_1}\cdots\phi_{i_n}\right]\\
	&=\int\mathcal{D}J\, \exp\left[{+\sum_{n=1}^{\infty}\frac{(-i)^n}{n!}(B_n)_{i_1\cdots i_n}J^{i_1}\cdots J^{i_n}}\right]e^{i J^i \phi_i}\,.\label{DictIntForm3}
\end{align}

Comparing \eqref{DictIntForm} and \eqref{DictIntForm3}, we see that a syntactic replacement \footnote{With an alternative wavefunction sign convention $\hat{\rho}_2\equiv -\rho_2>0$, the duality mapping reads $g:\hat{\rho}_2\mapsto B_2$ while the $n\neq 2$ mapping rules remain the same.}
\begin{align}
		g: \left(\begin{aligned}
		\rho_n\\B_n
		\end{aligned}\right)\mapsto (-i)^n\left(\begin{aligned}
			B_n\\\rho_n\label{dualityTransform}
			\end{aligned}\right)\,,
\end{align}
maps them to each other (after some labelling). Since both \eqref{DictIntForm} and \eqref{DictIntForm3} are equivalent statements of the correlator-wavefunction dictionary, we deduce that $g$ is an exact symmetry of the dictionary. Since $g^4=1$, the duality mapping generates a $\mathbb{Z}_4$ group that maps the dictionary to itself. The formal proof above shows that such a $\mathbb{Z}_4$ symmetry is valid to \textit{arbitrary} finite orders in perturbation theory, since the Gaussian integrals in both \eqref{DictIntForm} and \eqref{DictIntForm3} are well-defined as long as $\rho_2<0<B_2$. Note that in this derivation we have dropped an integration constant since ultimately the relationship between the $\rho_n$ and $B_n$ comes from taking derivatives with respect to an auxiliary current and therefore an overall constant in \eqref{DictIntForm3} is inconsequential.

To demonstrate the duality, let us inspect some simple examples. Up to $n=5$ at tree-level, we have
\begin{subequations}
	\begin{align}
		B_2&=-\frac{1}{\rho_2}~,\\
		B_3&=-\frac{1}{\rho_2^3}\,\rho_3~,\\
		B_4&=\frac{1}{\rho_2^4}\left[\rho_4-\left(\rho_3\frac{1}{\rho_2}\rho_3+\text{2 perms}\right)\right]~,\\
        \nonumber B_5&=-\frac{1}{\rho_2^5}\Bigg[\rho_5-\left(\rho_4\frac{1}{\rho_2}\rho_3+\text{9 perms}\right)\\
		&\qquad\qquad\quad+\left(\rho_3\frac{1}{\rho_2}\rho_3\frac{1}{\rho_2}\rho_3+\text{14 perms}\right)\Bigg]\,,
	\end{align}\label{TreeDictionaryRhoToB}
\end{subequations} 
where the internal DeWitt indices are understood as contracted. Note that, up to minus signs, the coefficient of each term is unity thanks to the normalisation of the wavefunction. Under the duality mapping $g$, equations \eqref{TreeDictionaryRhoToB} become
\begin{subequations}
	\begin{align}
		-\rho_2&=-\frac{1}{-B_2}~,\\
		i \rho_3&=-\frac{1}{-B_2^3}\,i B_3~,\\
		\rho_4&=\frac{1}{B_2^4}\left[B_4-\left(i B_3\,\frac{1}{-B_2}\, i B_3+\text{2 perms}\right)\right]~,\\
        \nonumber -i\rho_5&=-\frac{1}{-B_2^5}\Bigg[-i B_5-\left(B_4\frac{1}{-B_2}i B_3+\text{9 perms}\right)\\
		&\quad+\left(i B_3\frac{1}{-B_2}i B_3\frac{1}{-B_2}i B_3+\text{14 perms}\right)\Bigg],
	\end{align}\label{TreeDictionaryBToRho}
\end{subequations}
 which are equivalent to solving the original equations \eqref{TreeDictionaryRhoToB} for the $\rho$'s. Note also that since the $n=1$ entry starts out at loop-level, we have neglected it here. In practice, it is convenient to remove the tree-level tadpoles, so that the Gaussian term dominates the typical field fluctuations. However, we note that the duality \eqref{dualityTransform} works for any values of $\rho_1$ and $B_1$, since convergence is always guaranteed by the Gaussian term at non-typically large field fluctuations. In the Supplemental Material, we explicitly verify the duality including the $n=1$ tadpole terms up to 4-point 1-loop order, and show that consistently keeping tadpole terms is essential for the duality to work. We have further successfully confirmed the validity of the duality with a channel-insensitive check at 9-point 4-loop order using a computer algorithm \footnote{The Mathematica notebook that verifies the duality is available \href{https://github.com/XTCosmo/Correlator-Wavefunction-Duality}{here}.}.

In general, the tree-level dictionary translating physical wavefunction coefficients to correlators (in the absence of linear mixings which we assume throughout) reads
\begin{align}
	B_n=\frac{1}{(-\rho_2)^n}\,\sum_{k=0}^{n-3}(-1)^k\Big(\text{$k$-cuts}\Big)_\rho~,
\end{align}
with
\begin{align}
	\nonumber\Big(\text{$k$-cuts}\Big)_\rho\equiv& \sum_{n-k\geq n_1\cdots n_{k+1}\geq 3} \Bigg[\,\rho_{n_1}\frac{1}{\rho_2}\rho_{n_2}\cdots \rho_{n_k}\frac{1}{\rho_2}\rho_{n_{k+1}}\\
	&+\text{$(\pi_{n_1\cdots n_{k+1}}-1)$-perms}\,\Bigg]~.\label{rhokcutsStructure}
\end{align}
The correlator-wavefunction duality then implies the reciprocal formula
\begin{align}
	\rho_n=\frac{1}{B_2^n}\,\sum_{k=0}^{n-3} (-1)^k\Big(\text{\rm $k$-cuts}\Big)_B~, \label{RhotoBTree}
\end{align}
where $\Big(\text{\rm $k$-cuts}\Big)_B$ is obtained from \eqref{rhokcutsStructure} via a \textit{syntactic} substitution $\rho\to B$. Notice that here we have applied the tree-level topology to write $(-i)^{n_1+\cdots+n_{k+1}-n}=(-1)^k$.

In summary, the power of the duality allows us to directly invert the dictionary without the need of ever performing the algebraic inversion in practice, and all the combinatorics are automatically left intact.

\section{Factorisation of parity-odd correlators}
Let's now see how the duality we derived above can be put to good use. In \cite{Stefanyszyn:2023qov}, we derived a factorisation theorem for cosmological correlators of the inflaton and graviton which states that $n$-point functions of these states are factorised into lower-point objects if these observables are parity-odd (PO) \footnote{For the graviton, the reality and factorisation theorems of \cite{Stefanyszyn:2023qov} hold once we sum over the two helicities.}. This theorem allows for correlators arising from the exchange of additional states of \textit{any} mass and integer spin (in addition to contact diagrams), and relies on the following small set of mild assumptions:
\begin{itemize}
    \item Unitarity and locality
    \item The tree-level approximation 
    \item Bunch-Davies vacuum conditions
    \item IR-convergence of the nested time integrals that compute cosmological correlators
    \item Scale invariance of the interactions
\end{itemize}
An immediate consequence of the theorem is the absence of total-energy singularities in the PO sector of primordial perturbations which leads to a very nice distinction between the PO and parity-even (PE) sectors \footnote{This is a tree-level statement: total-energy singularities can arise at loop-level \cite{Lee:2023jby}.}. The theorem does not, however, state what objects such correlators factorise into. This is where the duality we have derived here comes to fruition. Indeed, the basis of the factorisation theorem of \cite{Stefanyszyn:2023qov} is a proof that for the PO sector we have $\rho^{\text{PO}}_n = 0$ (as long as the external states are the inflaton and/or the helicity-summed graviton, and any internal states that are produced during inflation and decay into these massless states are in the \textit{complementary series} of dS representations or the $SO(3)$ representations of \cite{Bordin:2018pca}). This follows from the fact that for the PO sector we have $\rho^{\text{PO}}_n = \psi_n(\bfk) - \psi^{*}_n(\bfk)$ i.e. it is the imaginary part of wavefunction coefficients that contribute to cosmological observables, yet under the above assumptions, wavefunction coefficients are purely real. This can be proven on very general grounds by performing Wick rotations of the time variables and using the fact that the time-ordered part of the bulk-bulk propagator is purely real after this rotation \footnote{Note that with this rotation we are still computing dS wavefunction coefficients rather than EAdS correlators. For connections between dS and AdS observables see \cite{Sleight:2020obc,Sleight:2021plv}.}, as are the vertices and the bulk-boundary propagator \cite{Stefanyszyn:2023qov}. The result holds for exchanging fields of arbitrary integer spin \footnote{See \cite{Goodhew:2024eup} for a more formal perspective.}.

Turning our attention to \eqref{RhotoBTree}, in the PO sector the factorisation theorem of \cite{Stefanyszyn:2023qov} therefore implies
\begin{align}
	\frac{1}{B_2^n}\,\sum_{k=0}^{n-3} (-1)^k\left[\Big(\text{\rm $k$-cuts}\Big)_B\right]^{\rm PO}=\rho_n^{\rm PO} = 0\,,
\end{align}
which can be rearranged to yield a formula for $B^{\text{PO}}_n$ in terms of lower-point \textit{correlators}:
\begin{align}
	B^{\text{PO}}_n = \sum_{k=1}^{n-3} (-1)^{k-1}\left[\Big(\text{\rm $k$-cuts}\Big)_B\right]^{\rm PO}~,~\forall\, n\geq 4 \,. \label{CCF}
\end{align}
In these expressions $[\ldots]^{\text{PO}}$ indicates that we are projecting correlators onto their PO part, which for correlators of the inflaton means that we take the imaginary part \footnote{This follows from the fact that in momentum space the inflaton is a parity transformation away from being real.}. Note that this formula holds regardless of how the parity violation arises which could be due to parity-violating vertices or due to the exchange of a spinning field with a parity-violating two-point function. We therefore see that the correlator-wavefunction duality has enabled us to derive, using the factorisation theorem of \cite{Stefanyszyn:2023qov}, a correlator-to-correlator factorisation (CCF) formula for the PO sector of primordial perturbations which has a neat interpretation in terms of correlator cuts. As an example, for $n=4$ there is only one possibility where an exchange diagram is cut into two cubic diagrams and \eqref{CCF} can be diagrammatically represented by
\begin{align}
   &\left(\raisebox{-10pt}{\includegraphics[scale=0.8]{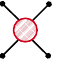}}\right)_{\text{tree}}^{\text{PO}}=~3\left(\raisebox{-10pt}{\includegraphics[scale=0.8]{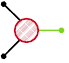}}\cdot\frac{1}{\raisebox{-4 pt}{\includegraphics[scale=0.8]{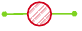}}}\cdot \raisebox{-10pt}{\includegraphics[scale=0.8]{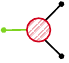}}\right)_{\text{tree}}^{\text{PO}} \,,
\end{align}
with the factor of $3$ symbolising the three different channels. The absence of a contact diagram contribution was already noted in \cite{Liu:2019fag,Cabass:2022rhr} \footnote{This resembles the situation for parity-odd $\text{Weyl}^3$ gravity \cite{Maldacena:2011nz,Shiraishi:2011st}. Indeed, parity-violation in the gravitational sector is also hard to come by \cite{Creminelli:2014wna,Cabass:2021fnw,Bordin:2020eui,Bartolo:2017szm}.}. For $n=5$ there is more structure with single and double cuts possible. We have
\begin{align}
   &\left(\raisebox{-10pt}{\includegraphics[scale=0.8]{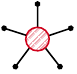}}\right)_{\text{tree}}^{\text{PO}}=~10\left(\raisebox{-10pt}{\includegraphics[scale=0.8]{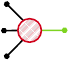}}\cdot\frac{1}{\raisebox{-4 pt}{\includegraphics[scale=0.8]{B_2g}}}\cdot \raisebox{-10pt}{\includegraphics[scale=0.8]{B_3_Rg}}\right)_{\text{tree}}^{\text{PO}}\nonumber\\
   &~-15\left(\raisebox{-10pt}{\includegraphics[scale=0.8]{B_3_L}}\cdot\frac{1}{\raisebox{-4 pt}{\includegraphics[scale=0.8]{B_2g}}}\cdot \raisebox{-4pt}{\includegraphics[scale=0.8]{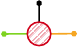}}\cdot\frac{1}{\raisebox{-4 pt}{\includegraphics[scale=0.8]{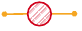}}}\cdot \raisebox{-10pt}{\includegraphics[scale=0.8]{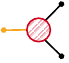}}\right)_{\text{tree}}^{\text{PO}}\,,
\end{align}
where again the numerical factors are counting distinct channels, and the different colours allow for the exchange of different fields.

In the above derivation we took additional states to be in the complementary series as this allowed us to set $\rho^{\text{PO}}_n = 0$ \cite{Stefanyszyn:2023qov}. If there are also principle series fields then this is not possible since then $\rho^{\text{PO}}_n \neq 0$. Correlators are indeed still factorised, but they do not factorise into other correlators. Our CCF formula is still useful for principle series fields, however, since if we have an explicit expression for $B^{\text{PO}}_{n}$ due to the exchange of complementary series fields (which can be computed using the CCF formula), we can analytically continue the mass parameter to derive corresponding expressions for the exchange of principle series fields \footnote{We need to take the $\eta_0 \rightarrow 0$ limit before performing any analytical continuation. We emphasise that the continuation is only possible at the level of the correlators and not wavefunction coefficients, since only the Schwinger-Keldysh propagators are analytic in the mass parameters. Indeed, the future boundary condition placed on the wavefunction bulk-bulk propagator makes a distinction between complementary and principle series fields.}.

The most relevant case for phenomenology is $n=4$ (corresponding to the trispectrum) and with curvature perturbations as the external states. This observable has received much attention recently due to the purported detection of parity-violation in the galaxy four-point function \cite{Hou:2022wfj,Philcox:2022hkh} (see also \cite{Philcox:2023ypl,Paul:2024uim}, however). In this case our CCF formula reads
\begin{align}
	B^{\text{PO}}_4 (\{ \bfk \}) =  &\left[ B_3(\bfk_1, \bfk_2, -\bfs) \cdot\frac{1}{B_{2}(\bfs)}\cdot B_3(\bfs, \bfk_3, \bfk_4) \right]^{\text{PO}} \nonumber \\
 & + (\bft + \bfu)\text{-channels} \,,
\end{align}
which provides a neat connection between distinct observables, namely between the PO part of the trispectrum of curvature perturbations, the bispectrum consisting of two curvature perturbations and one additional state with integer spin and mass in the complementary series, and the power spectrum of this new state. This relation is in principle testable, and any violation of this relation would imply one of the above listed assumptions is violated. One case of particular interest is where the exchanged field is the graviton. Parity-violation in such a set-up can come from a number of sources e.g. as a dynamical Chern-Simons correction to the graviton propagator \cite{Creque-Sarbinowski:2023wmb} or due to a mixing between the graviton and an $SU(2)$ gauge field \cite{Maleknejad:2016qjz,Maleknejad:2018nxz}. It would be interesting to study these cases in more detail given our CCF formula. 

Finally, let us point out that although the CCF relation \eqref{CCF} is almost completely transparent in the wavefunction language thanks the correlator-wavefunction duality, it seems rather mysterious from the traditional in-in/Schwinger-Keldysh (SK) diagrammatics of \cite{Chen:2017ryl}. One can indeed derive CCF relations using SK diagrammatics for specific lower-point examples, as we demonstrate for $n=4$ in the Supplemental Material, yet the general proof seems to be hidden from sight. It remains an intriguing question as to why boundary correlators should factorise into other correlators even away from specific kinematic limits (e.g. the OPE limit).

\section{Unitarity and cosmological correlators}
We now turn our attention to unitarity. Understanding the consequences of unitarity on observables is a vital component of a bootstrap toolbox. For scattering amplitudes unitarity requires tree-level processes to factorise near poles, and consistent factorisation across multiple channels can heavily constrain the space of admissible amplitudes and therefore admissible theories \cite{Benincasa:2007xk,Pajer:2020wnj}. More generally, unitarity imposes a set of Cutkosky cutting rules for scattering amplitudes \cite{Cutkosky:1960sp}. For cosmology the constraints imposed by unitarity are best understood at the level of wavefunction coefficients where a set of conditions imposed by unitary time evolution impose relations between analytically continued wavefunction coefficients \cite{Goodhew:2020hob,Melville:2021lst,Goodhew:2021oqg,Cespedes:2020xqq,Baumann:2021fxj}. Given that correlators rather than wavefunction coefficients are the true observables that are probed by cosmological surveys, it is desirable to find conditions that unitarity imposes directly on correlators. Progress in this direction has been made in e.g. \cite{Donath:2024utn}, however the general relations are not yet known. In this section we show how the duality we have derived in this paper can be used to convert the COT for wavefunction coefficients into relations between analytically continued cosmological correlators, and as an example we derive the general relation for the tree-level five-point function of massless scalars fields, with IR-finite interactions, in de Sitter space. To complement the other discussions in this work, we will assume parity-even interactions.

The above assumptions are useful since they imply that the wavefunction coefficients are purely real \cite{Stefanyszyn:2023qov,Cabass:2022rhr}. We therefore have $\rho_n = 2 \psi_n$ which we can use to wirte the COT in terms of $\rho_n$ followed by using our duality to convert the rules into statements about $B_n$. Let us illustrate this procedure for $n=4$ where unitarity imposes the following relations between $\rho_4$ and $\rho_3$ \cite{Goodhew:2020hob}:
\begin{align}
    2\,\text{Disc}_{s}\left(\raisebox{-8pt}{\includegraphics[scale=0.65]{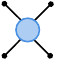}}\right)=\frac{1}{\raisebox{-3 pt}{\includegraphics[scale=0.65]{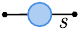}}}\text{Disc}_{s}\left(\raisebox{-8pt}{\includegraphics[scale=0.65]{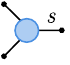}}\right)\times\text{Disc}_s\left( \raisebox{-8pt}{\includegraphics[scale=0.65]{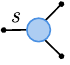}}\right), \label{FourPointCOT}
\end{align}
where $s=|\mathbf{k}_1+\mathbf{{k}}_2|$, and
\begin{align} \label{Disc}
\text{Disc}_{y} f(\{x \}, y) =f( \{ x\}, y) - f(\{ x \}, -y) \,,
\end{align}
with $\{ x \}$ a set of variables that do not flip sign under the discontinuity. Note that we do not include a complex conjugation on the right-hand side of \eqref{Disc}, in contrast to e.g. \cite{Goodhew:2021oqg}, since the wavefunction coefficients are real, and we have suppressed the dependence on spatial momenta and only included the dependence on the energies \footnote{This distinction requires an analytical continuation to non-physical momenta.}. By acting on the fully symmetric $\rho_4$ with $\text{Disc}_{s}$ we project onto the $s$-channel exchange diagram only. The relation \eqref{FourPointCOT} can be derived from the assumption of real couplings and factorisation properties of the bulk-bulk propagator \cite{Goodhew:2020hob}. We can now use the duality, namely the  relations between the $\rho_n$ and $B_n$ in \eqref{TreeDictionaryBToRho} on both sides of \eqref{FourPointCOT} to yield
\begin{align}
    &2\,\text{Disc}_{s}\left(\raisebox{-8pt}{\includegraphics[scale=0.65]{B_4}}\right)=\frac{1}{\raisebox{-3 pt}{\includegraphics[scale=0.65]{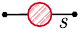}}}\text{Disc}_{s}\left(\raisebox{-8pt}{\includegraphics[scale=0.65]{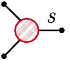}}\right)\times\text{Disc}_s\left( \raisebox{-8pt}{\includegraphics[scale=0.65]{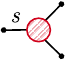}}\right),\label{FourPointBCOT}
\end{align}
where we have used $B_{2}(-s) = -B_{2}(s)$ which holds for massless fields. This recovers the expression derived in \cite{Goodhew:2020hob}. We emphasise that although this relation shares some similarities with the CCF formula, it requires us to work with non-physical momenta whereas the CCF formula is a true statement also for physical momenta. We can follow the same procedure for $n=5$ where unitarity imposes the following relation between $\rho_5$, $\rho_4$ and $\rho_3$:
\begin{align}
&2\,\text{Disc}_{s}\left(\raisebox{-8pt}{\includegraphics[scale=0.65]{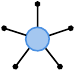}}\right)=\frac{1}{\raisebox{-3 pt}{\includegraphics[scale=0.65]{rho2_COT.pdf}}}\text{Disc}_{s}\left(\raisebox{-8pt}{\includegraphics[scale=0.65]{rho3_COT_L}}\right)\times\text{Disc}_s\left( \raisebox{-8pt}{\includegraphics[scale=0.65]{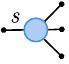}}\right).\label{FivePointCOT}
\end{align}
If we now use the relations \eqref{TreeDictionaryBToRho} on both sides of \eqref{FivePointCOT} we arrive at
\begin{align}
&2\,\text{Disc}_{s}\left(\raisebox{-8pt}{\includegraphics[scale=0.65]{B_5}}\right)=\frac{1}{\raisebox{-3 pt}{\includegraphics[scale=0.65]{B2_COT.pdf}}}\text{Disc}_{s}\left(\raisebox{-8pt}{\includegraphics[scale=0.65]{B3_COT_L}}\right)\times\text{Disc}_s\left( \raisebox{-8pt}{\includegraphics[scale=0.65]{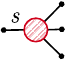}}\right)\label{FivePointBCOT},
\end{align}
which relies on a very non-trivial cancellation between the contributions with three copies of $B_3$. We emphasise that we have not restricted to particular diagrammatics or channels in the above relations: they hold for the full $\{\rho_n \}$ and $\{ B_n \}$ at tree-level, with the duality automatically keeping track of the combinatorics. We find it very interesting that the form of the COT is universal i.e. the $\rho$ relations are identical to the $B$ ones. This seems to suggest that the COT commutes with a naive application of the duality. We believe that this deserves further attention and we plan to return to it in the future. 
\section{Summary}

In this work we have derived a duality in the dictionary between cosmological correlators $\{B_n\}$ and the physical part of wavefunction coefficients $\{\rho_n\}$ that is valid to all orders in perturbation theory. This duality allows us to derive a reciprocal formula that reconstructs $\{\rho_n\}$ from $\{B_n\}$ in a syntactic fashion. When combined with the results of \cite{Stefanyszyn:2023qov} which states that $\rho^{\text{PO}}_n = 0$ for massless scalar/graviton external states, and complementary series internal states, we obtain an infinite set of correlator-to-correlator factorisation (CCF) formulae. These relations state that $n$-point PO correlators at tree-level are factorised into structured combinations of lower-point correlators. These CCF relations involve observables defined for \textit{physical} kinematics and can therefore in principle be tested observationally. Any violation of these relations would directly point to the failure of the tree-level assumption, unitarity, locality, scale invariance or the Bunch-Davies vacuum. We showed how our CCF formulae can be understood in terms of diagrammatic cuts, and since taking these cuts does not require any analytical continuation, they serve as the first example to understand the general structure of cosmological observables in a physically accessible manner. 

In addition to deriving physically testable relations, we also showed how the duality can be used to map the COT for wavefunction coefficients, which follows from unitary time evolution, into statements about cosmological correlators. Intriguingly, the form of the COT remains the same before and after application of the duality. We believe that this observation deserves further attention. It would also be interesting to use the duality to derive the COT for all $n$, thereby generalising what we have focused on in this paper for $n=4,5$.

Our work certainly opens up many more avenues for future exploration. For instance, the $\mathbb{Z}_4$ symmetry goes beyond the context of cosmology all the way to connected Green functions and EFT Wilson coefficients in general QFTs. It would be interesting to see if one can make general statements there too. In addition, it would be interesting to extend our CCF formula (or something akin to CCF) to loop-level. Finally, it would be neat to find a full proof of the CCF formula directly using the Schwinger-Keldysh formalism, where only observables are involved from the get-go.

\vskip 5pt

\paragraph*{Acknowledgements} We thank Paolo Benincasa, Trevor Cheung, Zongzhe Du, Sadra Jazayeri, Eiichiro Komatsu, Hayden Lee, Ciaran McCulloch, Enrico Pajer, Ayngaran Thavanesan, Dong-Gang Wang, Yi Wang, Zhong-Zhi Xianyu and Masahide Yamaguchi for helpful discussions. D.S. is supported by a UKRI Stephen Hawking Fellowship [grant number EP/W005441/1] and a Nottingham Research Fellowship from the University of Nottingham. X.T. is supported by STFC consolidated grants ST/T000694/1 and ST/X000664/1. Y.Z. is supported by the IBS under the project code, IBS-R018-D3. X.T. thanks the University of Nottingham for kind hospitality. For the purpose of open access, the authors have applied a CC BY public copyright licence to any Author Accepted Manuscript version arising. 

\paragraph*{Data access statement} No new data were created or analysed during this study.

\newpage
\appendix
\section{The Duality at 4-point, 1-loop} \label{OneLoopCheck}
In this section, we present an explicit check of the correlator-wavefunction duality at the 4-point 1-loop-level.
To establish a consistent power-counting scheme in perturbation theory, we have in mind an underlying interacting theory of the form $\lambda \Phi^3$, with $\lambda\ll 1$ being an expansion parameter. We use $\xi$ to count the number of loops, namely the number of unknown momenta that are integrated over. These can come from \textit{quantum loop} corrections to the wavefunction coefficients, or from \textit{classical loops} that arise when  we perform the path integral to take us from wavefunction coefficients to correlators. Without loss of generality, we organise the perturbation series expansion in powers of $\lambda$ and $\xi$. As long as the duality works at order $\mathcal{O}(\lambda^p \xi^q)$ for \textit{any} $p,q\in\mathbb{N}$, we can claim the validity of the duality is independent of this choice of power-counting scheme. Notice also that $B_n$ and $\rho_n$ are always of the same order in $\lambda$ and $\xi$. To illustrate this counting and to explain our diagrammatic notation, consider the following diagrams:

\begin{subequations}
\noindent
\begin{minipage}[t]{0.46\linewidth}
\begin{align}
    \raisebox{-3pt}{\includegraphics[scale=0.7]{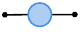}}&=\mathcal{O}(1)\,, \label{2ptFull}\\[8pt]
    \raisebox{-8pt}{\includegraphics[scale=0.7]{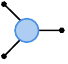}}&=\mathcal{O}(\lambda)\,, \label{3ptFull}\\[9pt]
     \raisebox{-3pt}{\includegraphics[scale=0.7]{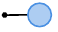}}&=\mathcal{O}(\lambda\xi)\,,\label{1ptFull} 
    \\\nonumber
\end{align}
\end{minipage}
\hfill
\begin{minipage}[t]{0.47\linewidth}
\begin{align} 
    \raisebox{-7pt}{\includegraphics[scale=0.7]{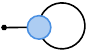}}&=\mathcal{O}(\lambda\xi)\,,\label{1ptCL} \\[0pt]
    \raisebox{-8pt}{\includegraphics[scale=0.7]{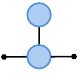}}&=\mathcal{O}(\lambda^2\xi)\,,\label{2ptFactorised}\\[5pt]
    \raisebox{-9pt}{\includegraphics[scale=0.7]{rho_4}}&=\mathcal{O}(\lambda^2)\,, \label{4ptFull}\
\end{align}
\end{minipage}
\end{subequations}
where we have removed the tree-level contribution to tadpoles and regarded both $\rho_1$ and $B_1$ as small loop-order perturbations. 
\eqref{2ptFull} simply represents $\rho_2$ computed to any order in perturbation theory. There is a tree-level contribution and therefore the leading contribution is independent of $\lambda$ and $\xi$. \eqref{3ptFull} and \eqref{1ptFull} similarly represent $\rho_3$ and $\rho_1$ computed to any order in perturbation theory. The former is $\mathcal{O}(\lambda)$ since the leading contribution comes from a tree-level contact diagram with a single cubic vertex, while for $\rho_1$ there is no tree-level contribution so the leading term is at one-loop and comes from a single cubic vertex and so it is $\mathcal{O}(\lambda \xi)$. \eqref{1ptCL} represents $\rho_3$ computed to any order in perturbation theory but with two of the external momenta fixed to be equal and opposite and integrated over thereby producing a contribution to the $1$-point function. Since $\rho_3$ starts out at tree-level and is $\mathcal{O}(\lambda)$, once we perform the momentum integration we have $\mathcal{O}(\lambda \xi)$. Note that if $\rho_3$ is computed at tree-level, \eqref{1ptCL} represents a ``classical loop" in the language of \cite{Lee:2023jby,Cespedes:2023aal}. \eqref{2ptFactorised} represents a product of $\rho_3$ and $\rho_1$ for which we still only have a single overall momentum-conserving delta function. Such contributions to correlation functions once we perform the Born rule are usually referred to as ``factorised contributions" which should be distinguished from disconnected contributions that carry more than one momentum-conserving delta function. Since $\rho_3$ starts at tree-level and is $\mathcal{O}(\lambda)$, while $\rho_1$ starts out at loop level and is $\mathcal{O}(\lambda \xi)$, this diagram is $\mathcal{O}(\lambda^2 \xi)$. Finally, \eqref{4ptFull} represents $\rho_4$ computed to any order in perturbation theory. The leading contribution comes from a tree-level exchange diagram so it is $\mathcal{O}(\lambda^2)$. 

As an explicit example we will work up to $\mathcal{O}(\lambda^2\xi)$, in which case the Born rule yields the following relations between the $B_n$ and $\rho_n$:
\begin{subequations}
    \begin{align}
    \raisebox{-4pt}{\includegraphics[scale=0.8]{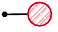}}~&=~-~\frac{\raisebox{-4pt}{\includegraphics[scale=0.8]{rho_1}}}{\raisebox{-6 pt}{\includegraphics[scale=0.8]{rho_2}}}~+~\frac{1}{2}~\frac{\raisebox{-4pt}{\includegraphics[scale=0.8]{rho_3_loop.pdf}}}{(\raisebox{-4pt}{\includegraphics[scale=0.8]{rho_2}})^2}~,\label{B1rho1}\\
    \raisebox{-4pt}{\includegraphics[scale=0.8]{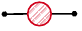}}~&=~-~\frac{1}{\raisebox{-6 pt}{\includegraphics[scale=0.8]{rho_2}}}~-~\frac{1}{2}~\frac{\raisebox{-4pt}{\includegraphics[scale=0.8]{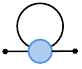}}}{\,~(\raisebox{-4pt}{\includegraphics[scale=0.8]{rho_2}})^3}~-~\frac{\raisebox{-4pt}{\includegraphics[scale=0.8]{rho3rho1.pdf}}}{~(\raisebox{-4pt}{\includegraphics[scale=0.8]{rho_2}})^3}\nonumber\\
    &+~\frac{1}{2}~\frac{\raisebox{-4pt}{\includegraphics[scale=0.8]{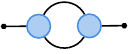}}}{\,~(\raisebox{-4pt}{\includegraphics[scale=0.8]{rho_2}})^4}~+~\frac{1}{2}~\frac{\raisebox{-4pt}{\includegraphics[scale=0.8]{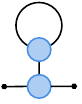}}}{~(\raisebox{-4pt}{\includegraphics[scale=0.8]{rho_2}})^4}~,\label{B2}\\
    \raisebox{-10pt}{\includegraphics[scale=0.8]{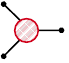}}~&=~-~\frac{\raisebox{-6 pt}{\includegraphics[scale=0.8]{rho_3}}}{(\raisebox{-4 pt}{\includegraphics[scale=0.8]{rho_2}})^3}~,\label{B3}\\
    \raisebox{-10pt}{\includegraphics[scale=0.8]{B_4}}~&=~\frac{\raisebox{-6 pt}{\includegraphics[scale=0.8]{rho_4}}}{~\,(\raisebox{-4 pt}{\includegraphics[scale=0.8]{rho_2}})^4}~-~3~\frac{\raisebox{-6 pt}{\includegraphics[scale=0.8]{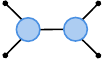}}}{~\,(\raisebox{-4 pt}{\includegraphics[scale=0.8]{rho_2}})^5}~.\label{B4}
\end{align}\label{4pt1loopDict}
\end{subequations} 
Note that the factor of $3$ in the second term on the RHS of \eqref{B4} is counting the three distinct channels whereas all other factors in other expressions are just numerical factors that arise from doing the path integral. Any relations that involve $B_n$ or $\rho_n$ with $n \geq 5$ are higher-order in $\lambda$ so we don't consider them here. As an example, to derive (\ref{B1rho1}), we first expand the generating functional \eqref{generatingFunctional} to $\mathcal{O}(J\lambda^2\xi)$,
\begin{align}
    Z[J]
    \nonumber&= \Det\left(\frac{-\rho_{2}}{2\pi}\right)^{-1/2}
    \Bigg\{1+i\,J^i\left(\rho_2^{-1}\right)_{ij}\Bigg[-(\rho_1)^j\\
    &\quad+\frac{1}{2}(\rho_3)^{jkl}\left(\rho_2^{-1}\right)_{kl}\Bigg]+\cdots\Bigg\}~,
\end{align}
and then differentiate to obtain the 1-point correlator
\begin{align}
    (B_1)_{i}&=\left.\frac{\partial}{i \,\partial J^{i}}\ln Z[J]\,\right|_{J=0}\nonumber\\
    &=-\left(\rho_2^{-1}\right)_{ij}(\rho_1)^j+\frac{1}{2}\left(\rho_2^{-1}\right)_{ij}(\rho_3)^{jkl}\left(\rho_2^{-1}\right)_{kl}~,
\end{align}
which corresponds to (\ref{B1rho1}). In momentum space for a single-field theory, this reads
\begin{align}
    B_1(\mathbf{k})=-\frac{\rho_1(\mathbf{k})}{\rho_2(\mathbf{k})}+\frac{1}{2}\int\frac{\d^3 q}{(2\pi)^3}\frac{\rho_3(\mathbf{k},\mathbf{q},-\mathbf{q})}{\rho_2(\mathbf{k})\rho_2(\mathbf{q})}~,
\end{align}
where all momentum-conserving $\delta$-functions are stripped. The second contribution on the RHS of this expression is the ``classical loop".  

To check the duality relation, we need to invert equations (\ref{B1rho1})-(\ref{B4}) and express $\rho$ in terms of $B$ before expanding and truncating at order $\mathcal{O}(\lambda^2\xi)$. The expressions for $\rho_3$ and $\rho_4$ can be easily worked out as
\begin{align}
    \raisebox{-10pt}{\includegraphics[scale=0.8]{rho_3}}~&=~\frac{\raisebox{-6 pt}{\includegraphics[scale=0.8]{B_3}}}{(\raisebox{-4 pt}{\includegraphics[scale=0.8]{B_2}})^3} \,,\label{rho3Rleation}\\
    \raisebox{-10pt}{\includegraphics[scale=0.8]{rho_4}}~&=~\frac{\raisebox{-6 pt}{\includegraphics[scale=0.8]{B_4}}}{~\,(\raisebox{-4 pt}{\includegraphics[scale=0.8]{B_2}})^4}~-~3~\frac{\raisebox{-6 pt}{\includegraphics[scale=0.8]{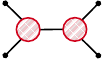}}}{~\,(\raisebox{-4 pt}{\includegraphics[scale=0.8]{B_2}})^5} \,.\label{rho4Rleation}
\end{align}
Up to $\mathcal{O}(\lambda^2 \xi)$, all contributions in \eqref{rho3Rleation} and \eqref{rho4Rleation} are in fact tree-level ones. With the expression \eqref{rho3Rleation} for $\rho_3$, we can then invert (\ref{B1rho1}) to get
\begin{align}
    \raisebox{-6pt}{\includegraphics[scale=0.8]{rho_1}}~=~\frac{\raisebox{-4pt}{\includegraphics[scale=0.8]{B_1}}}{\raisebox{-6 pt}{\includegraphics[scale=0.8]{B_2}}}~-~\frac{1}{2}~\frac{\raisebox{-4pt}{\includegraphics[scale=0.8]{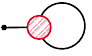}}}{(\raisebox{-4pt}{\includegraphics[scale=0.8]{B_2}})^2}~.\label{rho1Relation}
\end{align}
The $\rho_2$ case turns out to be more complicated. After solving for $\rho_2$ and substituting each $\rho_n$ by their expression in terms of $B_n$, $n=1,3,4$, we obtain
\begin{align}
    \raisebox{-4pt}{\includegraphics[scale=0.8]{rho_2}}~&=~-~\frac{1}{\raisebox{-6 pt}{\includegraphics[scale=0.8]{B_2}}}\nonumber\\
    &-\frac{1}{2}\frac{\raisebox{-4pt}{\includegraphics[scale=0.8]{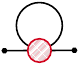}}}{\,~(\raisebox{-4pt}{\includegraphics[scale=0.8]{B_2}})^3}\,\textOrange{+\frac{2}{2}\,\frac{\raisebox{-4pt}{\includegraphics[scale=0.8]{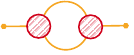}}}{\,~(\raisebox{-4pt}{\includegraphics[scale=0.8]{B_2o}})^4}}\,\textGreen{+\frac{1}{2}\,\frac{\raisebox{-4pt}{\includegraphics[scale=0.8]{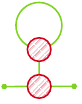}}}{\,~(\raisebox{-4pt}{\includegraphics[scale=0.8]{B_2g}})^4}}\nonumber\\
    &-\frac{\raisebox{-4pt}{\includegraphics[scale=0.8]{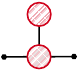}}}{~(\raisebox{-4pt}{\includegraphics[scale=0.8]{B_2}})^3}~\textGreen{+~\frac{1}{2}~\frac{\raisebox{-4pt}{\includegraphics[scale=0.8]{B3B3_tadG.pdf}}}{\,~(\raisebox{-4pt}{\includegraphics[scale=0.8]{B_2g}})^4}}\nonumber\\
    &\textOrange{-~\frac{1}{2}~\frac{\raisebox{-4pt}{\includegraphics[scale=0.8]{B3B3_loopO.pdf}}}{\,~(\raisebox{-4pt}{\includegraphics[scale=0.8]{B_2o}})^4}}\nonumber\\
    &\textGreen{-~\frac{1}{2}~\frac{\raisebox{-4pt}{\includegraphics[scale=0.8]{B3B3_tadG.pdf}}}{~(\raisebox{-4pt}{\includegraphics[scale=0.8]{B_2g}})^4}}~,
\end{align}
where the last four lines correspond to the last four terms in \eqref{B2}, respectively. Note that among the three channels of $B_3\cdot B_3$, two of them contract into a bubble diagram, and the remaining channel becomes a 1-loop tadpole diagram (the last term in the second line here). We have highlighted identical terms with the same colour to indicate the intricate cancellation of combinatorics at play here. Finally, we arrive at
\begin{align}
    \raisebox{-4pt}{\includegraphics[scale=0.8]{rho_2}}~&=~-~\frac{1}{\raisebox{-6 pt}{\includegraphics[scale=0.8]{B_2}}}~-~\frac{1}{2}~\frac{\raisebox{-4pt}{\includegraphics[scale=0.8]{B_4_loop.pdf}}}{\,~(\raisebox{-4pt}{\includegraphics[scale=0.8]{B_2}})^3}~-~\frac{\raisebox{-4pt}{\includegraphics[scale=0.8]{B3B1.pdf}}}{~(\raisebox{-4pt}{\includegraphics[scale=0.8]{B_2}})^3}\nonumber\\
    &+~\frac{1}{2}~\frac{\raisebox{-4pt}{\includegraphics[scale=0.8]{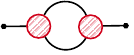}}}{\,~(\raisebox{-4pt}{\includegraphics[scale=0.8]{B_2}})^4}~+~\frac{1}{2}~\frac{\raisebox{-4pt}{\includegraphics[scale=0.8]{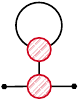}}}{~(\raisebox{-4pt}{\includegraphics[scale=0.8]{B_2}})^4}~. \label{rho2Rleation}
\end{align}
Now it is straightforward to check the duality. Applying $g:(\rho_n,B_n)\to (-i)^n (B_n,\rho_n)$ to \eqref{4pt1loopDict}, we indeed land on the desired equations \eqref{rho1Relation},\eqref{rho2Rleation},\eqref{rho3Rleation} and \eqref{rho4Rleation}.

As we can see from above, the inclusion of tadpole terms $\rho_1,B_1$ is essential in maintaining the duality. In practice, insisting tadpole cancellation $\rho_1=0$ or $B_1=0$ would then spontaneously break the duality at loop-level. However, the $\mathbb{Z}_4$ symmetry is still secretly at play, meaning the structure of the inverted dictionary is not arbitrary: it must be consistent with the $\mathbb{Z}_4$ symmetry if one reinserts the tadpole terms.

Before ending this section, we point out that the inversion procedure above can be automated by a computer algorithm. We have successfully confirmed the validity of the duality with a channel-insensitive check at 9-point 4-loop order.\footnote{See \href{https://github.com/XTCosmo/Correlator-Wavefunction-Duality}{here} for the Mathematica codes verifying the duality at any finite orders.}

\section{Correlator-to-Correlator Factorisation from Schwinger-Keldysh Diagrams at 4pt} \label{SKderivation}

Even though we have used the duality and the result that $\rho^{\text{PO}}_n = 0$ to prove that PO correlators are factorised into combinations of lower-point correlators for any $n$, it is instructive to see how we could draw the same conclusion directly at the level of Schwinger-Keldysh diagrams. We will illustrate this for the simplest case of $n=4$, and to be concrete we will focus on the PO trispectrum of curvature perturbations generated via the exchange of a massive spin-$1$ field (the generalisation to other theories is straightforward). In principle, $B^{\text{PO}}_4$ is computed as a sum over contact and exchange diagrams. However, it is now very well-known that the contact diagram contributions at tree-level yield a vanishing correlator once both sides of the in-in contour are added together \cite{Liu:2019fag,Cabass:2022rhr} (as is also the case for parity-odd $\text{Weyl}^3$ graviton correlator \cite{Maldacena:2011nz,Shiraishi:2011st} and the same cancellation explains the simplicity and rarity of other parity-odd graviton correlators \cite{Creminelli:2014wna,Cabass:2021fnw,Bordin:2020eui,Bartolo:2017szm}), so let's instead focus solely on exchange diagrams. We assume that there are two cubic vertices each of the form $\pi \pi \sigma$ where $\pi$ is the inflaton and $\sigma$ is the massive spin-$1$ field with its index suppressed. Such vertices can be constructed within the effective field theory of inflation (EFToI), as in \cite{Lee:2016vti}. The precise form of these interactions will play little role for us, however we take one vertex to be PO and the other to be PE such that we realise a PO trispectrum. Since cubic interactions of $SO(3)$ scalars cannot violate parity, only the transverse modes of the massive field will contribute. Diagrammatically, we therefore have \cite{Chen:2017ryl}
\begin{align}
    B_4~ & =  \raisebox{-15pt}{\includegraphics[scale=0.65]{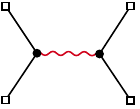}}~+~\raisebox{-15pt}{\includegraphics[scale=0.65]{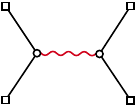}}\qquad\nonumber\\
    \nonumber\\
    &+ \raisebox{-15pt}{\includegraphics[scale=0.65]{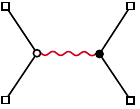}}~+~ \raisebox{-15pt}{\includegraphics[scale=0.65]{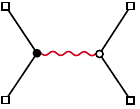}}~.\label{SKdiagrams}
\end{align}
The external propagators are given by
\begin{align}
K^{\pi}_{+} &= \pi(\eta_0,k) \pi^{*}(\eta,k)\,, \\
K^{\pi}_{-} &= (K^{\pi}_{+})^{*} \,,\label{PlustoMinusExternalProp}
\end{align}
while the internal ones are given by 
\begin{align}
G^{\sigma}_{++} &= \sigma(\eta,k)\sigma^{*}( \eta',k)\theta(\eta-\eta') + (\eta \leftrightarrow \eta')\,, \\
G^{\sigma}_{+-} &= \sigma^{*}(\eta,k)\sigma( \eta',k)\,, \\
G^{\sigma}_{--} &= (G^{\sigma}_{++})^{*}\,, \\
G^{\sigma}_{-+} &= (G^{\sigma}_{+-})^{*}\,,
\end{align}
where a $+/-$ indicates that a propagator is connected to a black/white vertex, and $\eta_0$ indicates the time at which the correlator is evaluated and we will take $\eta_0 \rightarrow 0$.\footnote{Note that the form of these propagators holds for any field. For example, if the massive field goes to the boundary we will have $K^{\sigma}_{+} = \sigma(\eta_0, k) \sigma^{*}(\eta, k)$.} Concentrating on the transverse modes only, for the massive spin-$1$ field we write $\sigma^{\pm}_i = \sigma \epsilon^{\pm}_i$ where $\epsilon^{\pm}_i$ are the transverse polarisation vectors and $\sigma$ is the mode function which is given by \cite{Lee:2016vti}
\begin{align}
\sigma(k, \eta) = \sqrt{\frac{\pi}{2}}\frac{1}{\sqrt{2 k }} e^{\frac{i \pi}{4} + \frac{i \pi \nu}{2}} (- k \eta)^{1/2} H_{\nu}^{(1)}(- k \eta) \,,
\end{align}
with 
\begin{align}
\nu  = \sqrt{\frac{1}{4} - \frac{m^2}{H^2}}\,.
\end{align}
Here $H_{\nu}^{(1)}$ is the Hankel function of the first kind. As always, in \eqref{SKdiagrams} we sum over the two helicity modes in each exchange diagram. Now, recall that black and white vertices are related by complex conjugation \textit{and} a flip of all momenta at that vertex. Since we are interested in the PO sector, flipping the sign of all momenta yields a minus sign so we can write
\begin{align}
    B^{\text{PO}}_4 = \left(\raisebox{-15pt}{\includegraphics[scale=0.65]{SKB4_bbv2}}~+~\raisebox{-15pt}{\includegraphics[scale=0.65]{SKB4_bwv2}}\right)~-~\text{c.c.}\,, \label{SKdiagramsNew}
\end{align}
which manifestly shows that PO correlators are purely imaginary. In \cite{Stefanyszyn:2023qov}, the Feynman propagator $G$ was written as the sum of two new propagators $G = C + F$ with all time-orderings contained within $C$ (the connected part indicated by double lines), and with the role of $F$ (the factorised part indicated by dashed lines) to ensure that $C$ is \textit{purely real} once both time variables are rotated clockwise by $90^{\circ}$ in the complex plane. For fields in the complementary series, which will  be our focus here and requires $m^2 < H^2 / 4$, $C$ in fact becomes the bulk-bulk propagator of the wavefunction formalism (for principle series fields the situation is different \cite{Stefanyszyn:2023qov}). We therefore have
\begin{align}
C^{\sigma}_{++} &= G^{\sigma}_{++} - \frac{\sigma(\eta_0,k)}{\sigma^{*}(\eta_0,k)}\sigma^{*}(\eta,k)\sigma^{*}(\eta',k)\,, \\
F^{\sigma}_{++} &= \frac{\sigma(\eta_0,k)}{\sigma^{*}(\eta_0,k)}\sigma^{*}(\eta,k)\sigma^{*}(\eta',k)\,.
\end{align}
With this decomposition, \eqref{SKdiagramsNew} becomes
\begin{align} \label{C-Fdecomposition}
    B^{\text{PO}}_4 &= \left(\raisebox{-15pt}{\includegraphics[scale=0.65]{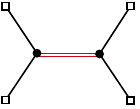}}+\raisebox{-15pt}{\includegraphics[scale=0.65]{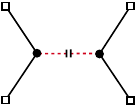}}+\raisebox{-15pt}{\includegraphics[scale=0.65]{SKB4_bwv2}}\right) \nonumber \\
   & -\text{c.c.}\,.
\end{align}
By construction, $C_{++}$ is purely real after a Wick rotation of both time variables to Euclidean time, while the other components that enter in the Feynman rules that compute the first diagram in \eqref{C-Fdecomposition} are also real after the same rotation \cite{Stefanyszyn:2023qov}:
\begin{align}
&\text{Integration measures}:  i \int \frac{d \eta}{\eta^4}, \quad i \int \frac{d \eta'}{\eta'^4} \,, \\
&\text{Derivatives}:  \eta \partial_{\eta}, \quad i \eta \bfk \,, \\
&\text{External propagator}:  {K}^{\pi}_{+} = \frac{H^2}{2 k^3}(1 - i k \eta)e^{i k \eta} \,, \\
&\text{Polarisations}: \sum_{h=\pm 1}\epsilon^{(h)}_{i}(\bfk) \epsilon^{(h) *}_{j}(\bfk) = \delta_{ij} - \hat{k}_{i}\hat{k}_{j} \,.
\label{PlusExternalProp}
\end{align}
Note that the derivatives take this form due to scale invariance. If the time integrals are convergent,\footnote{This is guaranteed if we work within the EFToI since the non-linear symmetries of the inflaton dictate a minimal number of derivatives that is sufficient to cancel the would-be divergences arising from the integration measure \cite{Stefanyszyn:2023qov}.} we can close the contour, drop the arc at infinity thanks to the Bunch-Davies vacuum choice, and use the Wick rotation to compute the necessary integrals. Since everything is manifestly real in Euclidean time, the diagrams with $C$ propagators cancel. This is a summary of one of the main results of \cite{Stefanyszyn:2023qov}. We therefore have 
\begin{align}
    B^{\text{PO}}_4 = \left(\raisebox{-15pt}{\includegraphics[scale=0.65]{SKB4_Fv2}}+\raisebox{-15pt}{\includegraphics[scale=0.65]{SKB4_bwv2}}\right)-\text{c.c.}\,. \label{FactorisedDiagram}
\end{align}
We now note that the $F^{\sigma}_{++}$ and $G^{\sigma}_{+-}$ propagators can be written as 
\begin{align}  
F^{\sigma}_{++} &= \frac{1}{B^{\sigma}_{2}(k)}K^{\sigma}_{+}(\eta, k)K^{\sigma}_{+}(\eta', k)\,, \\
G^{\sigma}_{+-} &= \frac{1}{B^{\sigma}_{2}(k)}K^{\sigma}_{+}(\eta, k)K^{\sigma}_{-}(\eta', k)\,,
\end{align}
where $B^{\sigma}_{2}$ is the power spectrum of $\sigma$. It follows that \eqref{FactorisedDiagram} can be expressed as
\begin{align}
    B^{\text{PO}}_4&=\frac{1}{~\raisebox{-2pt}{\includegraphics[scale=0.65]{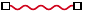}}} \left(\raisebox{-15pt}{\includegraphics[scale=0.65]{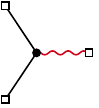}}~\raisebox{-15pt}{\includegraphics[scale=0.65]{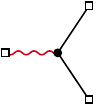}}+\raisebox{-15pt}{\includegraphics[scale=0.65]{SKB3bL_v2}}~\raisebox{-15pt}{\includegraphics[scale=0.65]{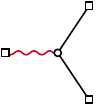}}\right.\nonumber\\
    &\qquad\qquad+\left.\raisebox{-15pt}{\includegraphics[scale=0.65]{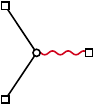}}~\raisebox{-15pt}{\includegraphics[scale=0.65]{SKB3wR_v2}}+\raisebox{-15pt}{\includegraphics[scale=0.65]{SKB3wL_v2}}~\raisebox{-15pt}{\includegraphics[scale=0.65]{SKB3bR_v2}}\right)\nonumber\\
    &=\left(\raisebox{-15pt}{\includegraphics[scale=0.65]{SKB3bL_v2}}+\raisebox{-15pt}{\includegraphics[scale=0.65]{SKB3wL_v2}}\right)\frac{1}{~\raisebox{-2pt}{\includegraphics[scale=0.65]{SKB2}}}\left(\raisebox{-15pt}{\includegraphics[scale=0.65]{SKB3bR_v2}}+\raisebox{-15pt}{\includegraphics[scale=0.65]{SKB3wR_v2}}\right)\nonumber\\
    &= \left[B_{3} \cdot \frac{1}{B_{2}} \cdot B_{3} \right]^{\text{PO}}\,, \nonumber \\
    & = \sum_{h = \pm 1} \left[ B^{h}_{3} \frac{1}{B_{2}^h} B_{3}^h \right]^{\text{PO}} \,,
\end{align}
where in the final line we have converted to the helicity basis with $B_{3}^{h}$ the three-point function with two curvature perturbations and the helicity-$h$ mode of the spin-$1$ field. We have also used that in the SK formalism three-point functions for $\pi \pi \sigma$ are computed via \cite{Chen:2017ryl}
\begin{align}
B_{3} = \raisebox{-15pt}{\includegraphics[scale=0.65]{SKB3bL_v2}}+\raisebox{-15pt}{\includegraphics[scale=0.65]{SKB3wL_v2}} \,.
\end{align}
We have therefore recovered the result we derived in the main text (there is no factor of $3$ here since we have worked with only one of the three channels). For principle series fields we can arrive at an equation similar to \eqref{FactorisedDiagram}, but the two diagrams do not have the same factors and therefore the RHS is not simply a combination of correlators. 




~\\~\\~\\~\\~\\~\\~\\~\\~\\~\\~\\~\\~\\~\\~\\~\\

\bibliography{Refs}

\begin{thebibliography}{104}%
\makeatletter
\providecommand \@ifxundefined [1]{%
 \@ifx{#1\undefined}
}%
\providecommand \@ifnum [1]{%
 \ifnum #1\expandafter \@firstoftwo
 \else \expandafter \@secondoftwo
 \fi
}%
\providecommand \@ifx [1]{%
 \ifx #1\expandafter \@firstoftwo
 \else \expandafter \@secondoftwo
 \fi
}%
\providecommand \natexlab [1]{#1}%
\providecommand \enquote  [1]{``#1''}%
\providecommand \bibnamefont  [1]{#1}%
\providecommand \bibfnamefont [1]{#1}%
\providecommand \citenamefont [1]{#1}%
\providecommand \href@noop [0]{\@secondoftwo}%
\providecommand \href [0]{\begingroup \@sanitize@url \@href}%
\providecommand \@href[1]{\@@startlink{#1}\@@href}%
\providecommand \@@href[1]{\endgroup#1\@@endlink}%
\providecommand \@sanitize@url [0]{\catcode `\\12\catcode `\$12\catcode
  `\&12\catcode `\#12\catcode `\^12\catcode `\_12\catcode `\%12\relax}%
\providecommand \@@startlink[1]{}%
\providecommand \@@endlink[0]{}%
\providecommand \url  [0]{\begingroup\@sanitize@url \@url }%
\providecommand \@url [1]{\endgroup\@href {#1}{\urlprefix }}%
\providecommand \urlprefix  [0]{URL }%
\providecommand \Eprint [0]{\href }%
\providecommand \doibase [0]{https://doi.org/}%
\providecommand \selectlanguage [0]{\@gobble}%
\providecommand \bibinfo  [0]{\@secondoftwo}%
\providecommand \bibfield  [0]{\@secondoftwo}%
\providecommand \translation [1]{[#1]}%
\providecommand \BibitemOpen [0]{}%
\providecommand \bibitemStop [0]{}%
\providecommand \bibitemNoStop [0]{.\EOS\space}%
\providecommand \EOS [0]{\spacefactor3000\relax}%
\providecommand \BibitemShut  [1]{\csname bibitem#1\endcsname}%
\let\auto@bib@innerbib\@empty
\bibitem [{\citenamefont {Chen}\ \emph {et~al.}(2016)\citenamefont {Chen},
  \citenamefont {Namjoo},\ and\ \citenamefont {Wang}}]{Chen:2015lza}%
  \BibitemOpen
  \bibfield  {author} {\bibinfo {author} {\bibfnamefont {X.}~\bibnamefont
  {Chen}}, \bibinfo {author} {\bibfnamefont {M.~H.}\ \bibnamefont {Namjoo}},\
  and\ \bibinfo {author} {\bibfnamefont {Y.}~\bibnamefont {Wang}},\ }\bibfield
  {title} {\bibinfo {title} {{Quantum Primordial Standard Clocks}},\ }\href
  {https://doi.org/10.1088/1475-7516/2016/02/013} {\bibfield  {journal}
  {\bibinfo  {journal} {JCAP}\ }\textbf {\bibinfo {volume} {02}},\ \bibinfo
  {pages} {013}},\ \Eprint {https://arxiv.org/abs/1509.03930} {arXiv:1509.03930
  [astro-ph.CO]} \BibitemShut {NoStop}%
\bibitem [{\citenamefont {Wang}\ \emph {et~al.}(2020)\citenamefont {Wang},
  \citenamefont {Wang},\ and\ \citenamefont {Zhu}}]{Wang:2020aqc}%
  \BibitemOpen
  \bibfield  {author} {\bibinfo {author} {\bibfnamefont {Y.}~\bibnamefont
  {Wang}}, \bibinfo {author} {\bibfnamefont {Z.}~\bibnamefont {Wang}},\ and\
  \bibinfo {author} {\bibfnamefont {Y.}~\bibnamefont {Zhu}},\ }\bibfield
  {title} {\bibinfo {title} {{Non-standard primordial clocks from induced mass
  in alternative to inflation scenarios}},\ }\href
  {https://doi.org/10.1088/1475-7516/2020/11/026} {\bibfield  {journal}
  {\bibinfo  {journal} {JCAP}\ }\textbf {\bibinfo {volume} {11}},\ \bibinfo
  {pages} {026}},\ \Eprint {https://arxiv.org/abs/2007.09677} {arXiv:2007.09677
  [hep-th]} \BibitemShut {NoStop}%
\bibitem [{\citenamefont {Chen}\ and\ \citenamefont
  {Wang}(2010)}]{Chen:2009zp}%
  \BibitemOpen
  \bibfield  {author} {\bibinfo {author} {\bibfnamefont {X.}~\bibnamefont
  {Chen}}\ and\ \bibinfo {author} {\bibfnamefont {Y.}~\bibnamefont {Wang}},\
  }\bibfield  {title} {\bibinfo {title} {{Quasi-Single Field Inflation and
  Non-Gaussianities}},\ }\href {https://doi.org/10.1088/1475-7516/2010/04/027}
  {\bibfield  {journal} {\bibinfo  {journal} {JCAP}\ }\textbf {\bibinfo
  {volume} {04}},\ \bibinfo {pages} {027}},\ \Eprint
  {https://arxiv.org/abs/0911.3380} {arXiv:0911.3380 [hep-th]} \BibitemShut
  {NoStop}%
\bibitem [{\citenamefont {Baumann}\ and\ \citenamefont
  {Green}(2012)}]{Baumann:2011nk}%
  \BibitemOpen
  \bibfield  {author} {\bibinfo {author} {\bibfnamefont {D.}~\bibnamefont
  {Baumann}}\ and\ \bibinfo {author} {\bibfnamefont {D.}~\bibnamefont
  {Green}},\ }\bibfield  {title} {\bibinfo {title} {{Signatures of
  Supersymmetry from the Early Universe}},\ }\href
  {https://doi.org/10.1103/PhysRevD.85.103520} {\bibfield  {journal} {\bibinfo
  {journal} {Phys. Rev. D}\ }\textbf {\bibinfo {volume} {85}},\ \bibinfo
  {pages} {103520} (\bibinfo {year} {2012})},\ \Eprint
  {https://arxiv.org/abs/1109.0292} {arXiv:1109.0292 [hep-th]} \BibitemShut
  {NoStop}%
\bibitem [{\citenamefont {Noumi}\ \emph {et~al.}(2013)\citenamefont {Noumi},
  \citenamefont {Yamaguchi},\ and\ \citenamefont {Yokoyama}}]{Noumi:2012vr}%
  \BibitemOpen
  \bibfield  {author} {\bibinfo {author} {\bibfnamefont {T.}~\bibnamefont
  {Noumi}}, \bibinfo {author} {\bibfnamefont {M.}~\bibnamefont {Yamaguchi}},\
  and\ \bibinfo {author} {\bibfnamefont {D.}~\bibnamefont {Yokoyama}},\
  }\bibfield  {title} {\bibinfo {title} {{Effective field theory approach to
  quasi-single field inflation and effects of heavy fields}},\ }\href
  {https://doi.org/10.1007/JHEP06(2013)051} {\bibfield  {journal} {\bibinfo
  {journal} {JHEP}\ }\textbf {\bibinfo {volume} {06}},\ \bibinfo {pages}
  {051}},\ \Eprint {https://arxiv.org/abs/1211.1624} {arXiv:1211.1624 [hep-th]}
  \BibitemShut {NoStop}%
\bibitem [{\citenamefont {Arkani-Hamed}\ and\ \citenamefont
  {Maldacena}(2015)}]{Arkani-Hamed:2015bza}%
  \BibitemOpen
  \bibfield  {author} {\bibinfo {author} {\bibfnamefont {N.}~\bibnamefont
  {Arkani-Hamed}}\ and\ \bibinfo {author} {\bibfnamefont {J.}~\bibnamefont
  {Maldacena}},\ }\bibfield  {title} {\bibinfo {title} {{Cosmological Collider
  Physics}},\ }\href@noop {} {\  (\bibinfo {year} {2015})},\ \Eprint
  {https://arxiv.org/abs/1503.08043} {arXiv:1503.08043 [hep-th]} \BibitemShut
  {NoStop}%
\bibitem [{\citenamefont {Lee}\ \emph {et~al.}(2016)\citenamefont {Lee},
  \citenamefont {Baumann},\ and\ \citenamefont {Pimentel}}]{Lee:2016vti}%
  \BibitemOpen
  \bibfield  {author} {\bibinfo {author} {\bibfnamefont {H.}~\bibnamefont
  {Lee}}, \bibinfo {author} {\bibfnamefont {D.}~\bibnamefont {Baumann}},\ and\
  \bibinfo {author} {\bibfnamefont {G.~L.}\ \bibnamefont {Pimentel}},\
  }\bibfield  {title} {\bibinfo {title} {{Non-Gaussianity as a Particle
  Detector}},\ }\href {https://doi.org/10.1007/JHEP12(2016)040} {\bibfield
  {journal} {\bibinfo  {journal} {JHEP}\ }\textbf {\bibinfo {volume} {12}},\
  \bibinfo {pages} {040}},\ \Eprint {https://arxiv.org/abs/1607.03735}
  {arXiv:1607.03735 [hep-th]} \BibitemShut {NoStop}%
\bibitem [{\citenamefont {Jazayeri}\ and\ \citenamefont
  {Renaux-Petel}(2022)}]{Jazayeri:2022kjy}%
  \BibitemOpen
  \bibfield  {author} {\bibinfo {author} {\bibfnamefont {S.}~\bibnamefont
  {Jazayeri}}\ and\ \bibinfo {author} {\bibfnamefont {S.}~\bibnamefont
  {Renaux-Petel}},\ }\bibfield  {title} {\bibinfo {title} {{Cosmological
  bootstrap in slow motion}},\ }\href {https://doi.org/10.1007/JHEP12(2022)137}
  {\bibfield  {journal} {\bibinfo  {journal} {JHEP}\ }\textbf {\bibinfo
  {volume} {12}},\ \bibinfo {pages} {137}},\ \Eprint
  {https://arxiv.org/abs/2205.10340} {arXiv:2205.10340 [hep-th]} \BibitemShut
  {NoStop}%
\bibitem [{\citenamefont {Pimentel}\ and\ \citenamefont
  {Wang}(2022)}]{Pimentel:2022fsc}%
  \BibitemOpen
  \bibfield  {author} {\bibinfo {author} {\bibfnamefont {G.~L.}\ \bibnamefont
  {Pimentel}}\ and\ \bibinfo {author} {\bibfnamefont {D.-G.}\ \bibnamefont
  {Wang}},\ }\bibfield  {title} {\bibinfo {title} {{Boostless cosmological
  collider bootstrap}},\ }\href {https://doi.org/10.1007/JHEP10(2022)177}
  {\bibfield  {journal} {\bibinfo  {journal} {JHEP}\ }\textbf {\bibinfo
  {volume} {10}},\ \bibinfo {pages} {177}},\ \Eprint
  {https://arxiv.org/abs/2205.00013} {arXiv:2205.00013 [hep-th]} \BibitemShut
  {NoStop}%
\bibitem [{\citenamefont {Cabass}\ \emph {et~al.}(2024)\citenamefont {Cabass},
  \citenamefont {Philcox}, \citenamefont {Ivanov}, \citenamefont {Akitsu},
  \citenamefont {Chen}, \citenamefont {Simonovi\'c},\ and\ \citenamefont
  {Zaldarriaga}}]{Cabass:2024wob}%
  \BibitemOpen
  \bibfield  {author} {\bibinfo {author} {\bibfnamefont {G.}~\bibnamefont
  {Cabass}}, \bibinfo {author} {\bibfnamefont {O.~H.~E.}\ \bibnamefont
  {Philcox}}, \bibinfo {author} {\bibfnamefont {M.~M.}\ \bibnamefont {Ivanov}},
  \bibinfo {author} {\bibfnamefont {K.}~\bibnamefont {Akitsu}}, \bibinfo
  {author} {\bibfnamefont {S.-F.}\ \bibnamefont {Chen}}, \bibinfo {author}
  {\bibfnamefont {M.}~\bibnamefont {Simonovi\'c}},\ and\ \bibinfo {author}
  {\bibfnamefont {M.}~\bibnamefont {Zaldarriaga}},\ }\bibfield  {title}
  {\bibinfo {title} {{BOSS Constraints on Massive Particles during Inflation:
  The Cosmological Collider in Action}},\ }\href@noop {} {\  (\bibinfo {year}
  {2024})},\ \Eprint {https://arxiv.org/abs/2404.01894} {arXiv:2404.01894
  [astro-ph.CO]} \BibitemShut {NoStop}%
\bibitem [{\citenamefont {Chakraborty}\ and\ \citenamefont
  {Stout}(2024{\natexlab{a}})}]{Chakraborty:2023qbp}%
  \BibitemOpen
  \bibfield  {author} {\bibinfo {author} {\bibfnamefont {P.}~\bibnamefont
  {Chakraborty}}\ and\ \bibinfo {author} {\bibfnamefont {J.}~\bibnamefont
  {Stout}},\ }\bibfield  {title} {\bibinfo {title} {{Light scalars at the
  cosmological collider}},\ }\href {https://doi.org/10.1007/JHEP02(2024)021}
  {\bibfield  {journal} {\bibinfo  {journal} {JHEP}\ }\textbf {\bibinfo
  {volume} {02}},\ \bibinfo {pages} {021}},\ \Eprint
  {https://arxiv.org/abs/2310.01494} {arXiv:2310.01494 [hep-th]} \BibitemShut
  {NoStop}%
\bibitem [{\citenamefont {Chen}\ \emph {et~al.}(2022)\citenamefont {Chen},
  \citenamefont {Ebadi},\ and\ \citenamefont {Kumar}}]{Chen:2022vzh}%
  \BibitemOpen
  \bibfield  {author} {\bibinfo {author} {\bibfnamefont {X.}~\bibnamefont
  {Chen}}, \bibinfo {author} {\bibfnamefont {R.}~\bibnamefont {Ebadi}},\ and\
  \bibinfo {author} {\bibfnamefont {S.}~\bibnamefont {Kumar}},\ }\bibfield
  {title} {\bibinfo {title} {{Classical cosmological collider physics and
  primordial features}},\ }\href
  {https://doi.org/10.1088/1475-7516/2022/08/083} {\bibfield  {journal}
  {\bibinfo  {journal} {JCAP}\ }\textbf {\bibinfo {volume} {08}},\ \bibinfo
  {pages} {083}},\ \Eprint {https://arxiv.org/abs/2205.01107} {arXiv:2205.01107
  [hep-ph]} \BibitemShut {NoStop}%
\bibitem [{\citenamefont {Tong}\ and\ \citenamefont
  {Xianyu}(2022)}]{Tong:2022cdz}%
  \BibitemOpen
  \bibfield  {author} {\bibinfo {author} {\bibfnamefont {X.}~\bibnamefont
  {Tong}}\ and\ \bibinfo {author} {\bibfnamefont {Z.-Z.}\ \bibnamefont
  {Xianyu}},\ }\bibfield  {title} {\bibinfo {title} {{Large spin-2 signals at
  the cosmological collider}},\ }\href
  {https://doi.org/10.1007/JHEP10(2022)194} {\bibfield  {journal} {\bibinfo
  {journal} {JHEP}\ }\textbf {\bibinfo {volume} {10}},\ \bibinfo {pages}
  {194}},\ \Eprint {https://arxiv.org/abs/2203.06349} {arXiv:2203.06349
  [hep-ph]} \BibitemShut {NoStop}%
\bibitem [{\citenamefont {Reece}\ \emph {et~al.}(2023)\citenamefont {Reece},
  \citenamefont {Wang},\ and\ \citenamefont {Xianyu}}]{Reece:2022soh}%
  \BibitemOpen
  \bibfield  {author} {\bibinfo {author} {\bibfnamefont {M.}~\bibnamefont
  {Reece}}, \bibinfo {author} {\bibfnamefont {L.-T.}\ \bibnamefont {Wang}},\
  and\ \bibinfo {author} {\bibfnamefont {Z.-Z.}\ \bibnamefont {Xianyu}},\
  }\bibfield  {title} {\bibinfo {title} {{Large-field inflation and the
  cosmological collider}},\ }\href
  {https://doi.org/10.1103/PhysRevD.107.L101304} {\bibfield  {journal}
  {\bibinfo  {journal} {Phys. Rev. D}\ }\textbf {\bibinfo {volume} {107}},\
  \bibinfo {pages} {L101304} (\bibinfo {year} {2023})},\ \Eprint
  {https://arxiv.org/abs/2204.11869} {arXiv:2204.11869 [hep-ph]} \BibitemShut
  {NoStop}%
\bibitem [{\citenamefont {Wang}\ and\ \citenamefont
  {Xianyu}(2020{\natexlab{a}})}]{Wang:2020ioa}%
  \BibitemOpen
  \bibfield  {author} {\bibinfo {author} {\bibfnamefont {L.-T.}\ \bibnamefont
  {Wang}}\ and\ \bibinfo {author} {\bibfnamefont {Z.-Z.}\ \bibnamefont
  {Xianyu}},\ }\bibfield  {title} {\bibinfo {title} {{Gauge Boson Signals at
  the Cosmological Collider}},\ }\href
  {https://doi.org/10.1007/JHEP11(2020)082} {\bibfield  {journal} {\bibinfo
  {journal} {JHEP}\ }\textbf {\bibinfo {volume} {11}},\ \bibinfo {pages}
  {082}},\ \Eprint {https://arxiv.org/abs/2004.02887} {arXiv:2004.02887
  [hep-ph]} \BibitemShut {NoStop}%
\bibitem [{\citenamefont {Kumar}\ and\ \citenamefont
  {Sundrum}(2020)}]{Kumar:2019ebj}%
  \BibitemOpen
  \bibfield  {author} {\bibinfo {author} {\bibfnamefont {S.}~\bibnamefont
  {Kumar}}\ and\ \bibinfo {author} {\bibfnamefont {R.}~\bibnamefont
  {Sundrum}},\ }\bibfield  {title} {\bibinfo {title} {{Cosmological Collider
  Physics and the Curvaton}},\ }\href {https://doi.org/10.1007/JHEP04(2020)077}
  {\bibfield  {journal} {\bibinfo  {journal} {JHEP}\ }\textbf {\bibinfo
  {volume} {04}},\ \bibinfo {pages} {077}},\ \Eprint
  {https://arxiv.org/abs/1908.11378} {arXiv:1908.11378 [hep-ph]} \BibitemShut
  {NoStop}%
\bibitem [{\citenamefont {Sohn}\ \emph {et~al.}(2024)\citenamefont {Sohn},
  \citenamefont {Wang}, \citenamefont {Fergusson},\ and\ \citenamefont
  {Shellard}}]{Sohn:2024xzd}%
  \BibitemOpen
  \bibfield  {author} {\bibinfo {author} {\bibfnamefont {W.}~\bibnamefont
  {Sohn}}, \bibinfo {author} {\bibfnamefont {D.-G.}\ \bibnamefont {Wang}},
  \bibinfo {author} {\bibfnamefont {J.~R.}\ \bibnamefont {Fergusson}},\ and\
  \bibinfo {author} {\bibfnamefont {E.~P.~S.}\ \bibnamefont {Shellard}},\
  }\bibfield  {title} {\bibinfo {title} {{Searching for Cosmological Collider
  in the Planck CMB Data}},\ }\href@noop {} {\  (\bibinfo {year} {2024})},\
  \Eprint {https://arxiv.org/abs/2404.07203} {arXiv:2404.07203 [astro-ph.CO]}
  \BibitemShut {NoStop}%
\bibitem [{\citenamefont {McCulloch}\ \emph {et~al.}(2024)\citenamefont
  {McCulloch}, \citenamefont {Pajer},\ and\ \citenamefont
  {Tong}}]{McCulloch:2024hiz}%
  \BibitemOpen
  \bibfield  {author} {\bibinfo {author} {\bibfnamefont {C.}~\bibnamefont
  {McCulloch}}, \bibinfo {author} {\bibfnamefont {E.}~\bibnamefont {Pajer}},\
  and\ \bibinfo {author} {\bibfnamefont {X.}~\bibnamefont {Tong}},\ }\bibfield
  {title} {\bibinfo {title} {{A cosmological tachyon collider: enhancing the
  long-short scale coupling}},\ }\href
  {https://doi.org/10.1007/JHEP05(2024)262} {\bibfield  {journal} {\bibinfo
  {journal} {JHEP}\ }\textbf {\bibinfo {volume} {05}},\ \bibinfo {pages}
  {262}},\ \Eprint {https://arxiv.org/abs/2401.11009} {arXiv:2401.11009
  [hep-th]} \BibitemShut {NoStop}%
\bibitem [{\citenamefont {Craig}\ \emph {et~al.}(2024)\citenamefont {Craig},
  \citenamefont {Kumar},\ and\ \citenamefont {McCune}}]{Craig:2024qgy}%
  \BibitemOpen
  \bibfield  {author} {\bibinfo {author} {\bibfnamefont {N.}~\bibnamefont
  {Craig}}, \bibinfo {author} {\bibfnamefont {S.}~\bibnamefont {Kumar}},\ and\
  \bibinfo {author} {\bibfnamefont {A.}~\bibnamefont {McCune}},\ }\bibfield
  {title} {\bibinfo {title} {{An Effective Cosmological Collider}},\
  }\href@noop {} {\  (\bibinfo {year} {2024})},\ \Eprint
  {https://arxiv.org/abs/2401.10976} {arXiv:2401.10976 [hep-ph]} \BibitemShut
  {NoStop}%
\bibitem [{\citenamefont {Jazayeri}\ \emph {et~al.}(2023)\citenamefont
  {Jazayeri}, \citenamefont {Renaux-Petel},\ and\ \citenamefont
  {Werth}}]{Jazayeri:2023xcj}%
  \BibitemOpen
  \bibfield  {author} {\bibinfo {author} {\bibfnamefont {S.}~\bibnamefont
  {Jazayeri}}, \bibinfo {author} {\bibfnamefont {S.}~\bibnamefont
  {Renaux-Petel}},\ and\ \bibinfo {author} {\bibfnamefont {D.}~\bibnamefont
  {Werth}},\ }\bibfield  {title} {\bibinfo {title} {{Shapes of the cosmological
  low-speed collider}},\ }\href {https://doi.org/10.1088/1475-7516/2023/12/035}
  {\bibfield  {journal} {\bibinfo  {journal} {JCAP}\ }\textbf {\bibinfo
  {volume} {12}},\ \bibinfo {pages} {035}},\ \Eprint
  {https://arxiv.org/abs/2307.01751} {arXiv:2307.01751 [hep-th]} \BibitemShut
  {NoStop}%
\bibitem [{\citenamefont {Qin}\ and\ \citenamefont
  {Xianyu}(2022)}]{Qin:2022lva}%
  \BibitemOpen
  \bibfield  {author} {\bibinfo {author} {\bibfnamefont {Z.}~\bibnamefont
  {Qin}}\ and\ \bibinfo {author} {\bibfnamefont {Z.-Z.}\ \bibnamefont
  {Xianyu}},\ }\bibfield  {title} {\bibinfo {title} {{Phase information in
  cosmological collider signals}},\ }\href
  {https://doi.org/10.1007/JHEP10(2022)192} {\bibfield  {journal} {\bibinfo
  {journal} {JHEP}\ }\textbf {\bibinfo {volume} {10}},\ \bibinfo {pages}
  {192}},\ \Eprint {https://arxiv.org/abs/2205.01692} {arXiv:2205.01692
  [hep-th]} \BibitemShut {NoStop}%
\bibitem [{\citenamefont {Bodas}\ \emph {et~al.}(2021)\citenamefont {Bodas},
  \citenamefont {Kumar},\ and\ \citenamefont {Sundrum}}]{Bodas:2020yho}%
  \BibitemOpen
  \bibfield  {author} {\bibinfo {author} {\bibfnamefont {A.}~\bibnamefont
  {Bodas}}, \bibinfo {author} {\bibfnamefont {S.}~\bibnamefont {Kumar}},\ and\
  \bibinfo {author} {\bibfnamefont {R.}~\bibnamefont {Sundrum}},\ }\bibfield
  {title} {\bibinfo {title} {{The Scalar Chemical Potential in Cosmological
  Collider Physics}},\ }\href {https://doi.org/10.1007/JHEP02(2021)079}
  {\bibfield  {journal} {\bibinfo  {journal} {JHEP}\ }\textbf {\bibinfo
  {volume} {02}},\ \bibinfo {pages} {079}},\ \Eprint
  {https://arxiv.org/abs/2010.04727} {arXiv:2010.04727 [hep-ph]} \BibitemShut
  {NoStop}%
\bibitem [{\citenamefont {Wang}\ and\ \citenamefont
  {Xianyu}(2020{\natexlab{b}})}]{Wang:2019gbi}%
  \BibitemOpen
  \bibfield  {author} {\bibinfo {author} {\bibfnamefont {L.-T.}\ \bibnamefont
  {Wang}}\ and\ \bibinfo {author} {\bibfnamefont {Z.-Z.}\ \bibnamefont
  {Xianyu}},\ }\bibfield  {title} {\bibinfo {title} {{In Search of Large
  Signals at the Cosmological Collider}},\ }\href
  {https://doi.org/10.1007/JHEP02(2020)044} {\bibfield  {journal} {\bibinfo
  {journal} {JHEP}\ }\textbf {\bibinfo {volume} {02}},\ \bibinfo {pages}
  {044}},\ \Eprint {https://arxiv.org/abs/1910.12876} {arXiv:1910.12876
  [hep-ph]} \BibitemShut {NoStop}%
\bibitem [{\citenamefont {Meerburg}\ \emph {et~al.}(2017)\citenamefont
  {Meerburg}, \citenamefont {M\"unchmeyer}, \citenamefont {Mu\~noz},\ and\
  \citenamefont {Chen}}]{Meerburg:2016zdz}%
  \BibitemOpen
  \bibfield  {author} {\bibinfo {author} {\bibfnamefont {P.~D.}\ \bibnamefont
  {Meerburg}}, \bibinfo {author} {\bibfnamefont {M.}~\bibnamefont
  {M\"unchmeyer}}, \bibinfo {author} {\bibfnamefont {J.~B.}\ \bibnamefont
  {Mu\~noz}},\ and\ \bibinfo {author} {\bibfnamefont {X.}~\bibnamefont
  {Chen}},\ }\bibfield  {title} {\bibinfo {title} {{Prospects for Cosmological
  Collider Physics}},\ }\href {https://doi.org/10.1088/1475-7516/2017/03/050}
  {\bibfield  {journal} {\bibinfo  {journal} {JCAP}\ }\textbf {\bibinfo
  {volume} {03}},\ \bibinfo {pages} {050}},\ \Eprint
  {https://arxiv.org/abs/1610.06559} {arXiv:1610.06559 [astro-ph.CO]}
  \BibitemShut {NoStop}%
\bibitem [{\citenamefont {Chen}\ \emph
  {et~al.}(2017{\natexlab{a}})\citenamefont {Chen}, \citenamefont {Wang},\ and\
  \citenamefont {Xianyu}}]{Chen:2016uwp}%
  \BibitemOpen
  \bibfield  {author} {\bibinfo {author} {\bibfnamefont {X.}~\bibnamefont
  {Chen}}, \bibinfo {author} {\bibfnamefont {Y.}~\bibnamefont {Wang}},\ and\
  \bibinfo {author} {\bibfnamefont {Z.-Z.}\ \bibnamefont {Xianyu}},\ }\bibfield
   {title} {\bibinfo {title} {{Standard Model Background of the Cosmological
  Collider}},\ }\href {https://doi.org/10.1103/PhysRevLett.118.261302}
  {\bibfield  {journal} {\bibinfo  {journal} {Phys. Rev. Lett.}\ }\textbf
  {\bibinfo {volume} {118}},\ \bibinfo {pages} {261302} (\bibinfo {year}
  {2017}{\natexlab{a}})},\ \Eprint {https://arxiv.org/abs/1610.06597}
  {arXiv:1610.06597 [hep-th]} \BibitemShut {NoStop}%
\bibitem [{\citenamefont {Bordin}\ \emph {et~al.}(2018)\citenamefont {Bordin},
  \citenamefont {Creminelli}, \citenamefont {Khmelnitsky},\ and\ \citenamefont
  {Senatore}}]{Bordin:2018pca}%
  \BibitemOpen
  \bibfield  {author} {\bibinfo {author} {\bibfnamefont {L.}~\bibnamefont
  {Bordin}}, \bibinfo {author} {\bibfnamefont {P.}~\bibnamefont {Creminelli}},
  \bibinfo {author} {\bibfnamefont {A.}~\bibnamefont {Khmelnitsky}},\ and\
  \bibinfo {author} {\bibfnamefont {L.}~\bibnamefont {Senatore}},\ }\bibfield
  {title} {\bibinfo {title} {{Light Particles with Spin in Inflation}},\ }\href
  {https://doi.org/10.1088/1475-7516/2018/10/013} {\bibfield  {journal}
  {\bibinfo  {journal} {JCAP}\ }\textbf {\bibinfo {volume} {10}},\ \bibinfo
  {pages} {013}},\ \Eprint {https://arxiv.org/abs/1806.10587} {arXiv:1806.10587
  [hep-th]} \BibitemShut {NoStop}%
\bibitem [{\citenamefont {Xianyu}\ and\ \citenamefont
  {Zang}(2024)}]{Xianyu:2023ytd}%
  \BibitemOpen
  \bibfield  {author} {\bibinfo {author} {\bibfnamefont {Z.-Z.}\ \bibnamefont
  {Xianyu}}\ and\ \bibinfo {author} {\bibfnamefont {J.}~\bibnamefont {Zang}},\
  }\bibfield  {title} {\bibinfo {title} {{Inflation correlators with multiple
  massive exchanges}},\ }\href {https://doi.org/10.1007/JHEP03(2024)070}
  {\bibfield  {journal} {\bibinfo  {journal} {JHEP}\ }\textbf {\bibinfo
  {volume} {03}},\ \bibinfo {pages} {070}},\ \Eprint
  {https://arxiv.org/abs/2309.10849} {arXiv:2309.10849 [hep-th]} \BibitemShut
  {NoStop}%
\bibitem [{\citenamefont {Aoki}\ \emph {et~al.}(2024)\citenamefont {Aoki},
  \citenamefont {Pinol}, \citenamefont {Sano}, \citenamefont {Yamaguchi},\ and\
  \citenamefont {Zhu}}]{Aoki:2024uyi}%
  \BibitemOpen
  \bibfield  {author} {\bibinfo {author} {\bibfnamefont {S.}~\bibnamefont
  {Aoki}}, \bibinfo {author} {\bibfnamefont {L.}~\bibnamefont {Pinol}},
  \bibinfo {author} {\bibfnamefont {F.}~\bibnamefont {Sano}}, \bibinfo {author}
  {\bibfnamefont {M.}~\bibnamefont {Yamaguchi}},\ and\ \bibinfo {author}
  {\bibfnamefont {Y.}~\bibnamefont {Zhu}},\ }\bibfield  {title} {\bibinfo
  {title} {{Cosmological Correlators with Double Massive Exchanges: Bootstrap
  Equation and Phenomenology}},\ }\href@noop {} {\  (\bibinfo {year} {2024})},\
  \Eprint {https://arxiv.org/abs/2404.09547} {arXiv:2404.09547 [hep-th]}
  \BibitemShut {NoStop}%
\bibitem [{\citenamefont {Chakraborty}\ and\ \citenamefont
  {Stout}(2024{\natexlab{b}})}]{Chakraborty:2023eoq}%
  \BibitemOpen
  \bibfield  {author} {\bibinfo {author} {\bibfnamefont {P.}~\bibnamefont
  {Chakraborty}}\ and\ \bibinfo {author} {\bibfnamefont {J.}~\bibnamefont
  {Stout}},\ }\bibfield  {title} {\bibinfo {title} {{Compact scalars at the
  cosmological collider}},\ }\href {https://doi.org/10.1007/JHEP03(2024)149}
  {\bibfield  {journal} {\bibinfo  {journal} {JHEP}\ }\textbf {\bibinfo
  {volume} {03}},\ \bibinfo {pages} {149}},\ \Eprint
  {https://arxiv.org/abs/2311.09219} {arXiv:2311.09219 [hep-th]} \BibitemShut
  {NoStop}%
\bibitem [{\citenamefont {Chen}\ \emph {et~al.}(2023)\citenamefont {Chen},
  \citenamefont {Fan},\ and\ \citenamefont {Li}}]{Chen:2023txq}%
  \BibitemOpen
  \bibfield  {author} {\bibinfo {author} {\bibfnamefont {X.}~\bibnamefont
  {Chen}}, \bibinfo {author} {\bibfnamefont {J.}~\bibnamefont {Fan}},\ and\
  \bibinfo {author} {\bibfnamefont {L.}~\bibnamefont {Li}},\ }\bibfield
  {title} {\bibinfo {title} {{New inflationary probes of axion dark matter}},\
  }\href {https://doi.org/10.1007/JHEP12(2023)197} {\bibfield  {journal}
  {\bibinfo  {journal} {JHEP}\ }\textbf {\bibinfo {volume} {12}},\ \bibinfo
  {pages} {197}},\ \Eprint {https://arxiv.org/abs/2303.03406} {arXiv:2303.03406
  [hep-ph]} \BibitemShut {NoStop}%
\bibitem [{\citenamefont {Cheung}\ \emph {et~al.}(2008)\citenamefont {Cheung},
  \citenamefont {Creminelli}, \citenamefont {Fitzpatrick}, \citenamefont
  {Kaplan},\ and\ \citenamefont {Senatore}}]{Cheung:2007st}%
  \BibitemOpen
  \bibfield  {author} {\bibinfo {author} {\bibfnamefont {C.}~\bibnamefont
  {Cheung}}, \bibinfo {author} {\bibfnamefont {P.}~\bibnamefont {Creminelli}},
  \bibinfo {author} {\bibfnamefont {A.~L.}\ \bibnamefont {Fitzpatrick}},
  \bibinfo {author} {\bibfnamefont {J.}~\bibnamefont {Kaplan}},\ and\ \bibinfo
  {author} {\bibfnamefont {L.}~\bibnamefont {Senatore}},\ }\bibfield  {title}
  {\bibinfo {title} {{The Effective Field Theory of Inflation}},\ }\href
  {https://doi.org/10.1088/1126-6708/2008/03/014} {\bibfield  {journal}
  {\bibinfo  {journal} {JHEP}\ }\textbf {\bibinfo {volume} {03}},\ \bibinfo
  {pages} {014}},\ \Eprint {https://arxiv.org/abs/0709.0293} {arXiv:0709.0293
  [hep-th]} \BibitemShut {NoStop}%
\bibitem [{\citenamefont {Chen}\ \emph {et~al.}(2007)\citenamefont {Chen},
  \citenamefont {Huang}, \citenamefont {Kachru},\ and\ \citenamefont
  {Shiu}}]{Chen:2006nt}%
  \BibitemOpen
  \bibfield  {author} {\bibinfo {author} {\bibfnamefont {X.}~\bibnamefont
  {Chen}}, \bibinfo {author} {\bibfnamefont {M.-x.}\ \bibnamefont {Huang}},
  \bibinfo {author} {\bibfnamefont {S.}~\bibnamefont {Kachru}},\ and\ \bibinfo
  {author} {\bibfnamefont {G.}~\bibnamefont {Shiu}},\ }\bibfield  {title}
  {\bibinfo {title} {{Observational signatures and non-Gaussianities of general
  single field inflation}},\ }\href
  {https://doi.org/10.1088/1475-7516/2007/01/002} {\bibfield  {journal}
  {\bibinfo  {journal} {JCAP}\ }\textbf {\bibinfo {volume} {01}},\ \bibinfo
  {pages} {002}},\ \Eprint {https://arxiv.org/abs/hep-th/0605045}
  {arXiv:hep-th/0605045} \BibitemShut {NoStop}%
\bibitem [{\citenamefont {Baumann}\ and\ \citenamefont
  {Green}(2011)}]{Baumann:2011su}%
  \BibitemOpen
  \bibfield  {author} {\bibinfo {author} {\bibfnamefont {D.}~\bibnamefont
  {Baumann}}\ and\ \bibinfo {author} {\bibfnamefont {D.}~\bibnamefont
  {Green}},\ }\bibfield  {title} {\bibinfo {title} {{Equilateral
  Non-Gaussianity and New Physics on the Horizon}},\ }\href
  {https://doi.org/10.1088/1475-7516/2011/09/014} {\bibfield  {journal}
  {\bibinfo  {journal} {JCAP}\ }\textbf {\bibinfo {volume} {09}},\ \bibinfo
  {pages} {014}},\ \Eprint {https://arxiv.org/abs/1102.5343} {arXiv:1102.5343
  [hep-th]} \BibitemShut {NoStop}%
\bibitem [{\citenamefont {Creminelli}(2003)}]{Creminelli:2003iq}%
  \BibitemOpen
  \bibfield  {author} {\bibinfo {author} {\bibfnamefont {P.}~\bibnamefont
  {Creminelli}},\ }\bibfield  {title} {\bibinfo {title} {{On non-Gaussianities
  in single-field inflation}},\ }\href
  {https://doi.org/10.1088/1475-7516/2003/10/003} {\bibfield  {journal}
  {\bibinfo  {journal} {JCAP}\ }\textbf {\bibinfo {volume} {10}},\ \bibinfo
  {pages} {003}},\ \Eprint {https://arxiv.org/abs/astro-ph/0306122}
  {arXiv:astro-ph/0306122} \BibitemShut {NoStop}%
\bibitem [{\citenamefont {Seery}\ and\ \citenamefont
  {Lidsey}(2005)}]{Seery:2005wm}%
  \BibitemOpen
  \bibfield  {author} {\bibinfo {author} {\bibfnamefont {D.}~\bibnamefont
  {Seery}}\ and\ \bibinfo {author} {\bibfnamefont {J.~E.}\ \bibnamefont
  {Lidsey}},\ }\bibfield  {title} {\bibinfo {title} {{Primordial
  non-Gaussianities in single field inflation}},\ }\href
  {https://doi.org/10.1088/1475-7516/2005/06/003} {\bibfield  {journal}
  {\bibinfo  {journal} {JCAP}\ }\textbf {\bibinfo {volume} {06}},\ \bibinfo
  {pages} {003}},\ \Eprint {https://arxiv.org/abs/astro-ph/0503692}
  {arXiv:astro-ph/0503692} \BibitemShut {NoStop}%
\bibitem [{\citenamefont {Holman}\ and\ \citenamefont
  {Tolley}(2008)}]{Holman:2007na}%
  \BibitemOpen
  \bibfield  {author} {\bibinfo {author} {\bibfnamefont {R.}~\bibnamefont
  {Holman}}\ and\ \bibinfo {author} {\bibfnamefont {A.~J.}\ \bibnamefont
  {Tolley}},\ }\bibfield  {title} {\bibinfo {title} {{Enhanced Non-Gaussianity
  from Excited Initial States}},\ }\href
  {https://doi.org/10.1088/1475-7516/2008/05/001} {\bibfield  {journal}
  {\bibinfo  {journal} {JCAP}\ }\textbf {\bibinfo {volume} {05}},\ \bibinfo
  {pages} {001}},\ \Eprint {https://arxiv.org/abs/0710.1302} {arXiv:0710.1302
  [hep-th]} \BibitemShut {NoStop}%
\bibitem [{\citenamefont {Meerburg}\ \emph {et~al.}(2009)\citenamefont
  {Meerburg}, \citenamefont {van~der Schaar},\ and\ \citenamefont
  {Corasaniti}}]{Meerburg:2009ys}%
  \BibitemOpen
  \bibfield  {author} {\bibinfo {author} {\bibfnamefont {P.~D.}\ \bibnamefont
  {Meerburg}}, \bibinfo {author} {\bibfnamefont {J.~P.}\ \bibnamefont {van~der
  Schaar}},\ and\ \bibinfo {author} {\bibfnamefont {P.~S.}\ \bibnamefont
  {Corasaniti}},\ }\bibfield  {title} {\bibinfo {title} {{Signatures of Initial
  State Modifications on Bispectrum Statistics}},\ }\href
  {https://doi.org/10.1088/1475-7516/2009/05/018} {\bibfield  {journal}
  {\bibinfo  {journal} {JCAP}\ }\textbf {\bibinfo {volume} {05}},\ \bibinfo
  {pages} {018}},\ \Eprint {https://arxiv.org/abs/0901.4044} {arXiv:0901.4044
  [hep-th]} \BibitemShut {NoStop}%
\bibitem [{\citenamefont {Ganc}(2011)}]{Ganc:2011dy}%
  \BibitemOpen
  \bibfield  {author} {\bibinfo {author} {\bibfnamefont {J.}~\bibnamefont
  {Ganc}},\ }\bibfield  {title} {\bibinfo {title} {{Calculating the local-type
  fNL for slow-roll inflation with a non-vacuum initial state}},\ }\href
  {https://doi.org/10.1103/PhysRevD.84.063514} {\bibfield  {journal} {\bibinfo
  {journal} {Phys. Rev. D}\ }\textbf {\bibinfo {volume} {84}},\ \bibinfo
  {pages} {063514} (\bibinfo {year} {2011})},\ \Eprint
  {https://arxiv.org/abs/1104.0244} {arXiv:1104.0244 [astro-ph.CO]}
  \BibitemShut {NoStop}%
\bibitem [{\citenamefont {Flauger}\ \emph {et~al.}(2013)\citenamefont
  {Flauger}, \citenamefont {Green},\ and\ \citenamefont
  {Porto}}]{Flauger:2013hra}%
  \BibitemOpen
  \bibfield  {author} {\bibinfo {author} {\bibfnamefont {R.}~\bibnamefont
  {Flauger}}, \bibinfo {author} {\bibfnamefont {D.}~\bibnamefont {Green}},\
  and\ \bibinfo {author} {\bibfnamefont {R.~A.}\ \bibnamefont {Porto}},\
  }\bibfield  {title} {\bibinfo {title} {{On squeezed limits in single-field
  inflation. Part I}},\ }\href {https://doi.org/10.1088/1475-7516/2013/08/032}
  {\bibfield  {journal} {\bibinfo  {journal} {JCAP}\ }\textbf {\bibinfo
  {volume} {08}},\ \bibinfo {pages} {032}},\ \Eprint
  {https://arxiv.org/abs/1303.1430} {arXiv:1303.1430 [hep-th]} \BibitemShut
  {NoStop}%
\bibitem [{\citenamefont {Green}\ and\ \citenamefont
  {Porto}(2020)}]{Green:2020whw}%
  \BibitemOpen
  \bibfield  {author} {\bibinfo {author} {\bibfnamefont {D.}~\bibnamefont
  {Green}}\ and\ \bibinfo {author} {\bibfnamefont {R.~A.}\ \bibnamefont
  {Porto}},\ }\bibfield  {title} {\bibinfo {title} {{Signals of a Quantum
  Universe}},\ }\href {https://doi.org/10.1103/PhysRevLett.124.251302}
  {\bibfield  {journal} {\bibinfo  {journal} {Phys. Rev. Lett.}\ }\textbf
  {\bibinfo {volume} {124}},\ \bibinfo {pages} {251302} (\bibinfo {year}
  {2020})},\ \Eprint {https://arxiv.org/abs/2001.09149} {arXiv:2001.09149
  [hep-th]} \BibitemShut {NoStop}%
\bibitem [{\citenamefont {Salcedo}\ \emph {et~al.}(2024)\citenamefont
  {Salcedo}, \citenamefont {Colas},\ and\ \citenamefont
  {Pajer}}]{Salcedo:2024smn}%
  \BibitemOpen
  \bibfield  {author} {\bibinfo {author} {\bibfnamefont {S.~A.}\ \bibnamefont
  {Salcedo}}, \bibinfo {author} {\bibfnamefont {T.}~\bibnamefont {Colas}},\
  and\ \bibinfo {author} {\bibfnamefont {E.}~\bibnamefont {Pajer}},\ }\bibfield
   {title} {\bibinfo {title} {{The Open Effective Field Theory of Inflation}},\
  }\href@noop {} {\  (\bibinfo {year} {2024})},\ \Eprint
  {https://arxiv.org/abs/2404.15416} {arXiv:2404.15416 [hep-th]} \BibitemShut
  {NoStop}%
\bibitem [{\citenamefont {Maldacena}(2003)}]{Maldacena:2002vr}%
  \BibitemOpen
  \bibfield  {author} {\bibinfo {author} {\bibfnamefont {J.~M.}\ \bibnamefont
  {Maldacena}},\ }\bibfield  {title} {\bibinfo {title} {{Non-Gaussian features
  of primordial fluctuations in single field inflationary models}},\ }\href
  {https://doi.org/10.1088/1126-6708/2003/05/013} {\bibfield  {journal}
  {\bibinfo  {journal} {JHEP}\ }\textbf {\bibinfo {volume} {05}},\ \bibinfo
  {pages} {013}},\ \Eprint {https://arxiv.org/abs/astro-ph/0210603}
  {arXiv:astro-ph/0210603} \BibitemShut {NoStop}%
\bibitem [{\citenamefont {Anninos}\ \emph {et~al.}(2015)\citenamefont
  {Anninos}, \citenamefont {Anous}, \citenamefont {Freedman},\ and\
  \citenamefont {Konstantinidis}}]{Anninos:2014lwa}%
  \BibitemOpen
  \bibfield  {author} {\bibinfo {author} {\bibfnamefont {D.}~\bibnamefont
  {Anninos}}, \bibinfo {author} {\bibfnamefont {T.}~\bibnamefont {Anous}},
  \bibinfo {author} {\bibfnamefont {D.~Z.}\ \bibnamefont {Freedman}},\ and\
  \bibinfo {author} {\bibfnamefont {G.}~\bibnamefont {Konstantinidis}},\
  }\bibfield  {title} {\bibinfo {title} {{Late-time Structure of the
  Bunch-Davies De Sitter Wavefunction}},\ }\href
  {https://doi.org/10.1088/1475-7516/2015/11/048} {\bibfield  {journal}
  {\bibinfo  {journal} {JCAP}\ }\textbf {\bibinfo {volume} {11}},\ \bibinfo
  {pages} {048}},\ \Eprint {https://arxiv.org/abs/1406.5490} {arXiv:1406.5490
  [hep-th]} \BibitemShut {NoStop}%
\bibitem [{\citenamefont {Arkani-Hamed}\ \emph {et~al.}(2020)\citenamefont
  {Arkani-Hamed}, \citenamefont {Baumann}, \citenamefont {Lee},\ and\
  \citenamefont {Pimentel}}]{Arkani-Hamed:2018kmz}%
  \BibitemOpen
  \bibfield  {author} {\bibinfo {author} {\bibfnamefont {N.}~\bibnamefont
  {Arkani-Hamed}}, \bibinfo {author} {\bibfnamefont {D.}~\bibnamefont
  {Baumann}}, \bibinfo {author} {\bibfnamefont {H.}~\bibnamefont {Lee}},\ and\
  \bibinfo {author} {\bibfnamefont {G.~L.}\ \bibnamefont {Pimentel}},\
  }\bibfield  {title} {\bibinfo {title} {{The Cosmological Bootstrap:
  Inflationary Correlators from Symmetries and Singularities}},\ }\href
  {https://doi.org/10.1007/JHEP04(2020)105} {\bibfield  {journal} {\bibinfo
  {journal} {JHEP}\ }\textbf {\bibinfo {volume} {04}},\ \bibinfo {pages}
  {105}},\ \Eprint {https://arxiv.org/abs/1811.00024} {arXiv:1811.00024
  [hep-th]} \BibitemShut {NoStop}%
\bibitem [{\citenamefont {Baumann}\ \emph {et~al.}(2020)\citenamefont
  {Baumann}, \citenamefont {Duaso~Pueyo}, \citenamefont {Joyce}, \citenamefont
  {Lee},\ and\ \citenamefont {Pimentel}}]{Baumann:2019oyu}%
  \BibitemOpen
  \bibfield  {author} {\bibinfo {author} {\bibfnamefont {D.}~\bibnamefont
  {Baumann}}, \bibinfo {author} {\bibfnamefont {C.}~\bibnamefont
  {Duaso~Pueyo}}, \bibinfo {author} {\bibfnamefont {A.}~\bibnamefont {Joyce}},
  \bibinfo {author} {\bibfnamefont {H.}~\bibnamefont {Lee}},\ and\ \bibinfo
  {author} {\bibfnamefont {G.~L.}\ \bibnamefont {Pimentel}},\ }\bibfield
  {title} {\bibinfo {title} {{The cosmological bootstrap: weight-shifting
  operators and scalar seeds}},\ }\href
  {https://doi.org/10.1007/JHEP12(2020)204} {\bibfield  {journal} {\bibinfo
  {journal} {JHEP}\ }\textbf {\bibinfo {volume} {12}},\ \bibinfo {pages}
  {204}},\ \Eprint {https://arxiv.org/abs/1910.14051} {arXiv:1910.14051
  [hep-th]} \BibitemShut {NoStop}%
\bibitem [{\citenamefont {Baumann}\ \emph {et~al.}(2021)\citenamefont
  {Baumann}, \citenamefont {Duaso~Pueyo}, \citenamefont {Joyce}, \citenamefont
  {Lee},\ and\ \citenamefont {Pimentel}}]{Baumann:2020dch}%
  \BibitemOpen
  \bibfield  {author} {\bibinfo {author} {\bibfnamefont {D.}~\bibnamefont
  {Baumann}}, \bibinfo {author} {\bibfnamefont {C.}~\bibnamefont
  {Duaso~Pueyo}}, \bibinfo {author} {\bibfnamefont {A.}~\bibnamefont {Joyce}},
  \bibinfo {author} {\bibfnamefont {H.}~\bibnamefont {Lee}},\ and\ \bibinfo
  {author} {\bibfnamefont {G.~L.}\ \bibnamefont {Pimentel}},\ }\bibfield
  {title} {\bibinfo {title} {{The Cosmological Bootstrap: Spinning Correlators
  from Symmetries and Factorization}},\ }\href
  {https://doi.org/10.21468/SciPostPhys.11.3.071} {\bibfield  {journal}
  {\bibinfo  {journal} {SciPost Phys.}\ }\textbf {\bibinfo {volume} {11}},\
  \bibinfo {pages} {071} (\bibinfo {year} {2021})},\ \Eprint
  {https://arxiv.org/abs/2005.04234} {arXiv:2005.04234 [hep-th]} \BibitemShut
  {NoStop}%
\bibitem [{\citenamefont {Baumann}\ \emph
  {et~al.}(2022{\natexlab{a}})\citenamefont {Baumann}, \citenamefont {Chen},
  \citenamefont {Duaso~Pueyo}, \citenamefont {Joyce}, \citenamefont {Lee},\
  and\ \citenamefont {Pimentel}}]{Baumann:2021fxj}%
  \BibitemOpen
  \bibfield  {author} {\bibinfo {author} {\bibfnamefont {D.}~\bibnamefont
  {Baumann}}, \bibinfo {author} {\bibfnamefont {W.-M.}\ \bibnamefont {Chen}},
  \bibinfo {author} {\bibfnamefont {C.}~\bibnamefont {Duaso~Pueyo}}, \bibinfo
  {author} {\bibfnamefont {A.}~\bibnamefont {Joyce}}, \bibinfo {author}
  {\bibfnamefont {H.}~\bibnamefont {Lee}},\ and\ \bibinfo {author}
  {\bibfnamefont {G.~L.}\ \bibnamefont {Pimentel}},\ }\bibfield  {title}
  {\bibinfo {title} {{Linking the singularities of cosmological correlators}},\
  }\href {https://doi.org/10.1007/JHEP09(2022)010} {\bibfield  {journal}
  {\bibinfo  {journal} {JHEP}\ }\textbf {\bibinfo {volume} {09}},\ \bibinfo
  {pages} {010}},\ \Eprint {https://arxiv.org/abs/2106.05294} {arXiv:2106.05294
  [hep-th]} \BibitemShut {NoStop}%
\bibitem [{\citenamefont {Albayrak}\ \emph {et~al.}(2023)\citenamefont
  {Albayrak}, \citenamefont {Benincasa},\ and\ \citenamefont
  {Duaso~Pueyo}}]{Albayrak:2023hie}%
  \BibitemOpen
  \bibfield  {author} {\bibinfo {author} {\bibfnamefont {S.}~\bibnamefont
  {Albayrak}}, \bibinfo {author} {\bibfnamefont {P.}~\bibnamefont
  {Benincasa}},\ and\ \bibinfo {author} {\bibfnamefont {C.}~\bibnamefont
  {Duaso~Pueyo}},\ }\bibfield  {title} {\bibinfo {title} {{Perturbative
  Unitarity and the Wavefunction of the Universe}},\ }\href@noop {} {\
  (\bibinfo {year} {2023})},\ \Eprint {https://arxiv.org/abs/2305.19686}
  {arXiv:2305.19686 [hep-th]} \BibitemShut {NoStop}%
\bibitem [{\citenamefont {Arkani-Hamed}\ \emph {et~al.}(2017)\citenamefont
  {Arkani-Hamed}, \citenamefont {Benincasa},\ and\ \citenamefont
  {Postnikov}}]{Arkani-Hamed:2017fdk}%
  \BibitemOpen
  \bibfield  {author} {\bibinfo {author} {\bibfnamefont {N.}~\bibnamefont
  {Arkani-Hamed}}, \bibinfo {author} {\bibfnamefont {P.}~\bibnamefont
  {Benincasa}},\ and\ \bibinfo {author} {\bibfnamefont {A.}~\bibnamefont
  {Postnikov}},\ }\bibfield  {title} {\bibinfo {title} {{Cosmological Polytopes
  and the Wavefunction of the Universe}},\ }\href@noop {} {\  (\bibinfo {year}
  {2017})},\ \Eprint {https://arxiv.org/abs/1709.02813} {arXiv:1709.02813
  [hep-th]} \BibitemShut {NoStop}%
\bibitem [{\citenamefont {Benincasa}(2022)}]{Benincasa:2022omn}%
  \BibitemOpen
  \bibfield  {author} {\bibinfo {author} {\bibfnamefont {P.}~\bibnamefont
  {Benincasa}},\ }\bibfield  {title} {\bibinfo {title}
  {{Wavefunctionals/S-matrix techniques in de Sitter}},\ }\href
  {https://doi.org/10.22323/1.406.0358} {\bibfield  {journal} {\bibinfo
  {journal} {PoS}\ }\textbf {\bibinfo {volume} {CORFU2021}},\ \bibinfo {pages}
  {358} (\bibinfo {year} {2022})},\ \Eprint {https://arxiv.org/abs/2203.16378}
  {arXiv:2203.16378 [hep-th]} \BibitemShut {NoStop}%
\bibitem [{\citenamefont {Goodhew}\ \emph
  {et~al.}(2021{\natexlab{a}})\citenamefont {Goodhew}, \citenamefont
  {Jazayeri},\ and\ \citenamefont {Pajer}}]{Goodhew:2020hob}%
  \BibitemOpen
  \bibfield  {author} {\bibinfo {author} {\bibfnamefont {H.}~\bibnamefont
  {Goodhew}}, \bibinfo {author} {\bibfnamefont {S.}~\bibnamefont {Jazayeri}},\
  and\ \bibinfo {author} {\bibfnamefont {E.}~\bibnamefont {Pajer}},\ }\bibfield
   {title} {\bibinfo {title} {{The Cosmological Optical Theorem}},\ }\href
  {https://doi.org/10.1088/1475-7516/2021/04/021} {\bibfield  {journal}
  {\bibinfo  {journal} {JCAP}\ }\textbf {\bibinfo {volume} {04}},\ \bibinfo
  {pages} {021}},\ \Eprint {https://arxiv.org/abs/2009.02898} {arXiv:2009.02898
  [hep-th]} \BibitemShut {NoStop}%
\bibitem [{\citenamefont {Melville}\ and\ \citenamefont
  {Pajer}(2021)}]{Melville:2021lst}%
  \BibitemOpen
  \bibfield  {author} {\bibinfo {author} {\bibfnamefont {S.}~\bibnamefont
  {Melville}}\ and\ \bibinfo {author} {\bibfnamefont {E.}~\bibnamefont
  {Pajer}},\ }\bibfield  {title} {\bibinfo {title} {{Cosmological Cutting
  Rules}},\ }\href {https://doi.org/10.1007/JHEP05(2021)249} {\bibfield
  {journal} {\bibinfo  {journal} {JHEP}\ }\textbf {\bibinfo {volume} {05}},\
  \bibinfo {pages} {249}},\ \Eprint {https://arxiv.org/abs/2103.09832}
  {arXiv:2103.09832 [hep-th]} \BibitemShut {NoStop}%
\bibitem [{\citenamefont {Goodhew}\ \emph
  {et~al.}(2021{\natexlab{b}})\citenamefont {Goodhew}, \citenamefont
  {Jazayeri}, \citenamefont {Lee},\ and\ \citenamefont
  {Pajer}}]{Goodhew:2021oqg}%
  \BibitemOpen
  \bibfield  {author} {\bibinfo {author} {\bibfnamefont {H.}~\bibnamefont
  {Goodhew}}, \bibinfo {author} {\bibfnamefont {S.}~\bibnamefont {Jazayeri}},
  \bibinfo {author} {\bibfnamefont {M.~H.~G.}\ \bibnamefont {Lee}},\ and\
  \bibinfo {author} {\bibfnamefont {E.}~\bibnamefont {Pajer}},\ }\bibfield
  {title} {\bibinfo {title} {{Cutting cosmological correlators}},\ }\href
  {https://doi.org/10.1088/1475-7516/2021/08/003} {\bibfield  {journal}
  {\bibinfo  {journal} {JCAP}\ }\textbf {\bibinfo {volume} {08}},\ \bibinfo
  {pages} {003}},\ \Eprint {https://arxiv.org/abs/2104.06587} {arXiv:2104.06587
  [hep-th]} \BibitemShut {NoStop}%
\bibitem [{\citenamefont {Pajer}(2021)}]{Pajer:2020wxk}%
  \BibitemOpen
  \bibfield  {author} {\bibinfo {author} {\bibfnamefont {E.}~\bibnamefont
  {Pajer}},\ }\bibfield  {title} {\bibinfo {title} {{Building a Boostless
  Bootstrap for the Bispectrum}},\ }\href
  {https://doi.org/10.1088/1475-7516/2021/01/023} {\bibfield  {journal}
  {\bibinfo  {journal} {JCAP}\ }\textbf {\bibinfo {volume} {01}},\ \bibinfo
  {pages} {023}},\ \Eprint {https://arxiv.org/abs/2010.12818} {arXiv:2010.12818
  [hep-th]} \BibitemShut {NoStop}%
\bibitem [{\citenamefont {Jazayeri}\ \emph {et~al.}(2021)\citenamefont
  {Jazayeri}, \citenamefont {Pajer},\ and\ \citenamefont
  {Stefanyszyn}}]{Jazayeri:2021fvk}%
  \BibitemOpen
  \bibfield  {author} {\bibinfo {author} {\bibfnamefont {S.}~\bibnamefont
  {Jazayeri}}, \bibinfo {author} {\bibfnamefont {E.}~\bibnamefont {Pajer}},\
  and\ \bibinfo {author} {\bibfnamefont {D.}~\bibnamefont {Stefanyszyn}},\
  }\bibfield  {title} {\bibinfo {title} {{From locality and unitarity to
  cosmological correlators}},\ }\href {https://doi.org/10.1007/JHEP10(2021)065}
  {\bibfield  {journal} {\bibinfo  {journal} {JHEP}\ }\textbf {\bibinfo
  {volume} {10}},\ \bibinfo {pages} {065}},\ \Eprint
  {https://arxiv.org/abs/2103.08649} {arXiv:2103.08649 [hep-th]} \BibitemShut
  {NoStop}%
\bibitem [{\citenamefont {Salcedo}\ \emph {et~al.}(2023)\citenamefont
  {Salcedo}, \citenamefont {Lee}, \citenamefont {Melville},\ and\ \citenamefont
  {Pajer}}]{Salcedo:2022aal}%
  \BibitemOpen
  \bibfield  {author} {\bibinfo {author} {\bibfnamefont {S.~A.}\ \bibnamefont
  {Salcedo}}, \bibinfo {author} {\bibfnamefont {M.~H.~G.}\ \bibnamefont {Lee}},
  \bibinfo {author} {\bibfnamefont {S.}~\bibnamefont {Melville}},\ and\
  \bibinfo {author} {\bibfnamefont {E.}~\bibnamefont {Pajer}},\ }\bibfield
  {title} {\bibinfo {title} {{The Analytic Wavefunction}},\ }\href
  {https://doi.org/10.1007/JHEP06(2023)020} {\bibfield  {journal} {\bibinfo
  {journal} {JHEP}\ }\textbf {\bibinfo {volume} {06}},\ \bibinfo {pages}
  {020}},\ \Eprint {https://arxiv.org/abs/2212.08009} {arXiv:2212.08009
  [hep-th]} \BibitemShut {NoStop}%
\bibitem [{\citenamefont {Meltzer}(2021)}]{Meltzer:2021zin}%
  \BibitemOpen
  \bibfield  {author} {\bibinfo {author} {\bibfnamefont {D.}~\bibnamefont
  {Meltzer}},\ }\bibfield  {title} {\bibinfo {title} {{The inflationary
  wavefunction from analyticity and factorization}},\ }\href
  {https://doi.org/10.1088/1475-7516/2021/12/018} {\bibfield  {journal}
  {\bibinfo  {journal} {JCAP}\ }\textbf {\bibinfo {volume} {12}}\bibfield
  {number} {\bibinfo  {number} { (12)},\ \bibinfo {pages} {018}},\ }\Eprint
  {https://arxiv.org/abs/2107.10266} {arXiv:2107.10266 [hep-th]} \BibitemShut
  {NoStop}%
\bibitem [{\citenamefont {Donath}\ and\ \citenamefont
  {Pajer}(2024)}]{Donath:2024utn}%
  \BibitemOpen
  \bibfield  {author} {\bibinfo {author} {\bibfnamefont {Y.}~\bibnamefont
  {Donath}}\ and\ \bibinfo {author} {\bibfnamefont {E.}~\bibnamefont {Pajer}},\
  }\bibfield  {title} {\bibinfo {title} {{The In-Out Formalism for In-In
  Correlators}},\ }\href@noop {} {\  (\bibinfo {year} {2024})},\ \Eprint
  {https://arxiv.org/abs/2402.05999} {arXiv:2402.05999 [hep-th]} \BibitemShut
  {NoStop}%
\bibitem [{\citenamefont {Lee}(2024)}]{Lee:2023kno}%
  \BibitemOpen
  \bibfield  {author} {\bibinfo {author} {\bibfnamefont {M.~H.~G.}\
  \bibnamefont {Lee}},\ }\bibfield  {title} {\bibinfo {title} {{From amplitudes
  to analytic wavefunctions}},\ }\href
  {https://doi.org/10.1007/JHEP03(2024)058} {\bibfield  {journal} {\bibinfo
  {journal} {JHEP}\ }\textbf {\bibinfo {volume} {03}},\ \bibinfo {pages}
  {058}},\ \Eprint {https://arxiv.org/abs/2310.01525} {arXiv:2310.01525
  [hep-th]} \BibitemShut {NoStop}%
\bibitem [{\citenamefont {Agui-Salcedo}\ and\ \citenamefont
  {Melville}(2023)}]{Agui-Salcedo:2023wlq}%
  \BibitemOpen
  \bibfield  {author} {\bibinfo {author} {\bibfnamefont {S.}~\bibnamefont
  {Agui-Salcedo}}\ and\ \bibinfo {author} {\bibfnamefont {S.}~\bibnamefont
  {Melville}},\ }\bibfield  {title} {\bibinfo {title} {{The cosmological tree
  theorem}},\ }\href {https://doi.org/10.1007/JHEP12(2023)076} {\bibfield
  {journal} {\bibinfo  {journal} {JHEP}\ }\textbf {\bibinfo {volume} {12}},\
  \bibinfo {pages} {076}},\ \Eprint {https://arxiv.org/abs/2308.00680}
  {arXiv:2308.00680 [hep-th]} \BibitemShut {NoStop}%
\bibitem [{\citenamefont {C\'espedes}\ \emph {et~al.}(2021)\citenamefont
  {C\'espedes}, \citenamefont {Davis},\ and\ \citenamefont
  {Melville}}]{Cespedes:2020xqq}%
  \BibitemOpen
  \bibfield  {author} {\bibinfo {author} {\bibfnamefont {S.}~\bibnamefont
  {C\'espedes}}, \bibinfo {author} {\bibfnamefont {A.-C.}\ \bibnamefont
  {Davis}},\ and\ \bibinfo {author} {\bibfnamefont {S.}~\bibnamefont
  {Melville}},\ }\bibfield  {title} {\bibinfo {title} {{On the time evolution
  of cosmological correlators}},\ }\href
  {https://doi.org/10.1007/JHEP02(2021)012} {\bibfield  {journal} {\bibinfo
  {journal} {JHEP}\ }\textbf {\bibinfo {volume} {02}},\ \bibinfo {pages}
  {012}},\ \Eprint {https://arxiv.org/abs/2009.07874} {arXiv:2009.07874
  [hep-th]} \BibitemShut {NoStop}%
\bibitem [{\citenamefont {Baumann}\ \emph
  {et~al.}(2022{\natexlab{b}})\citenamefont {Baumann}, \citenamefont {Green},
  \citenamefont {Joyce}, \citenamefont {Pajer}, \citenamefont {Pimentel},
  \citenamefont {Sleight},\ and\ \citenamefont {Taronna}}]{Baumann:2022jpr}%
  \BibitemOpen
  \bibfield  {author} {\bibinfo {author} {\bibfnamefont {D.}~\bibnamefont
  {Baumann}}, \bibinfo {author} {\bibfnamefont {D.}~\bibnamefont {Green}},
  \bibinfo {author} {\bibfnamefont {A.}~\bibnamefont {Joyce}}, \bibinfo
  {author} {\bibfnamefont {E.}~\bibnamefont {Pajer}}, \bibinfo {author}
  {\bibfnamefont {G.~L.}\ \bibnamefont {Pimentel}}, \bibinfo {author}
  {\bibfnamefont {C.}~\bibnamefont {Sleight}},\ and\ \bibinfo {author}
  {\bibfnamefont {M.}~\bibnamefont {Taronna}},\ }\bibfield  {title} {\bibinfo
  {title} {{Snowmass White Paper: The Cosmological Bootstrap}},\ }in\
  \href@noop {} {\emph {\bibinfo {booktitle} {{Snowmass 2021}}}}\ (\bibinfo
  {year} {2022})\ \Eprint {https://arxiv.org/abs/2203.08121} {arXiv:2203.08121
  [hep-th]} \BibitemShut {NoStop}%
\bibitem [{\citenamefont {Di~Pietro}\ \emph {et~al.}(2022)\citenamefont
  {Di~Pietro}, \citenamefont {Gorbenko},\ and\ \citenamefont
  {Komatsu}}]{DiPietro:2021sjt}%
  \BibitemOpen
  \bibfield  {author} {\bibinfo {author} {\bibfnamefont {L.}~\bibnamefont
  {Di~Pietro}}, \bibinfo {author} {\bibfnamefont {V.}~\bibnamefont
  {Gorbenko}},\ and\ \bibinfo {author} {\bibfnamefont {S.}~\bibnamefont
  {Komatsu}},\ }\bibfield  {title} {\bibinfo {title} {{Analyticity and
  unitarity for cosmological correlators}},\ }\href
  {https://doi.org/10.1007/JHEP03(2022)023} {\bibfield  {journal} {\bibinfo
  {journal} {JHEP}\ }\textbf {\bibinfo {volume} {03}},\ \bibinfo {pages}
  {023}},\ \Eprint {https://arxiv.org/abs/2108.01695} {arXiv:2108.01695
  [hep-th]} \BibitemShut {NoStop}%
\bibitem [{\citenamefont {Hogervorst}\ \emph {et~al.}(2023)\citenamefont
  {Hogervorst}, \citenamefont {Penedones},\ and\ \citenamefont
  {Vaziri}}]{Hogervorst:2021uvp}%
  \BibitemOpen
  \bibfield  {author} {\bibinfo {author} {\bibfnamefont {M.}~\bibnamefont
  {Hogervorst}}, \bibinfo {author} {\bibfnamefont {J.~a.}\ \bibnamefont
  {Penedones}},\ and\ \bibinfo {author} {\bibfnamefont {K.~S.}\ \bibnamefont
  {Vaziri}},\ }\bibfield  {title} {\bibinfo {title} {{Towards the
  non-perturbative cosmological bootstrap}},\ }\href
  {https://doi.org/10.1007/JHEP02(2023)162} {\bibfield  {journal} {\bibinfo
  {journal} {JHEP}\ }\textbf {\bibinfo {volume} {02}},\ \bibinfo {pages}
  {162}},\ \Eprint {https://arxiv.org/abs/2107.13871} {arXiv:2107.13871
  [hep-th]} \BibitemShut {NoStop}%
\bibitem [{\citenamefont {Melville}\ and\ \citenamefont
  {Pimentel}(2023)}]{Melville:2023kgd}%
  \BibitemOpen
  \bibfield  {author} {\bibinfo {author} {\bibfnamefont {S.}~\bibnamefont
  {Melville}}\ and\ \bibinfo {author} {\bibfnamefont {G.~L.}\ \bibnamefont
  {Pimentel}},\ }\bibfield  {title} {\bibinfo {title} {{A de Sitter $S$-matrix
  for the masses}},\ }\href@noop {} {\  (\bibinfo {year} {2023})},\ \Eprint
  {https://arxiv.org/abs/2309.07092} {arXiv:2309.07092 [hep-th]} \BibitemShut
  {NoStop}%
\bibitem [{\citenamefont {Melville}\ and\ \citenamefont
  {Pimentel}(2024)}]{Melville:2024ove}%
  \BibitemOpen
  \bibfield  {author} {\bibinfo {author} {\bibfnamefont {S.}~\bibnamefont
  {Melville}}\ and\ \bibinfo {author} {\bibfnamefont {G.~L.}\ \bibnamefont
  {Pimentel}},\ }\bibfield  {title} {\bibinfo {title} {{A de Sitter S-matrix
  from amputated cosmological correlators}},\ }\href@noop {} {\  (\bibinfo
  {year} {2024})},\ \Eprint {https://arxiv.org/abs/2404.05712}
  {arXiv:2404.05712 [hep-th]} \BibitemShut {NoStop}%
\bibitem [{\citenamefont {Fan}\ and\ \citenamefont
  {Xianyu}(2024)}]{Fan:2024iek}%
  \BibitemOpen
  \bibfield  {author} {\bibinfo {author} {\bibfnamefont {B.}~\bibnamefont
  {Fan}}\ and\ \bibinfo {author} {\bibfnamefont {Z.-Z.}\ \bibnamefont
  {Xianyu}},\ }\bibfield  {title} {\bibinfo {title} {{Cosmological Amplitudes
  in Power-Law FRW Universe}},\ }\href@noop {} {\  (\bibinfo {year} {2024})},\
  \Eprint {https://arxiv.org/abs/2403.07050} {arXiv:2403.07050 [hep-th]}
  \BibitemShut {NoStop}%
\bibitem [{\citenamefont {Gomez}\ \emph {et~al.}(2022)\citenamefont {Gomez},
  \citenamefont {Lipinski~Jusinskas},\ and\ \citenamefont
  {Lipstein}}]{Gomez:2021ujt}%
  \BibitemOpen
  \bibfield  {author} {\bibinfo {author} {\bibfnamefont {H.}~\bibnamefont
  {Gomez}}, \bibinfo {author} {\bibfnamefont {R.}~\bibnamefont
  {Lipinski~Jusinskas}},\ and\ \bibinfo {author} {\bibfnamefont
  {A.}~\bibnamefont {Lipstein}},\ }\bibfield  {title} {\bibinfo {title}
  {{Cosmological scattering equations at tree-level and one-loop}},\ }\href
  {https://doi.org/10.1007/JHEP07(2022)004} {\bibfield  {journal} {\bibinfo
  {journal} {JHEP}\ }\textbf {\bibinfo {volume} {07}},\ \bibinfo {pages}
  {004}},\ \Eprint {https://arxiv.org/abs/2112.12695} {arXiv:2112.12695
  [hep-th]} \BibitemShut {NoStop}%
\bibitem [{\citenamefont {C\'espedes}\ \emph {et~al.}(2024)\citenamefont
  {C\'espedes}, \citenamefont {Davis},\ and\ \citenamefont
  {Wang}}]{Cespedes:2023aal}%
  \BibitemOpen
  \bibfield  {author} {\bibinfo {author} {\bibfnamefont {S.}~\bibnamefont
  {C\'espedes}}, \bibinfo {author} {\bibfnamefont {A.-C.}\ \bibnamefont
  {Davis}},\ and\ \bibinfo {author} {\bibfnamefont {D.-G.}\ \bibnamefont
  {Wang}},\ }\bibfield  {title} {\bibinfo {title} {{On the IR divergences in de
  Sitter space: loops, resummation and the semi-classical wavefunction}},\
  }\href {https://doi.org/10.1007/JHEP04(2024)004} {\bibfield  {journal}
  {\bibinfo  {journal} {JHEP}\ }\textbf {\bibinfo {volume} {04}},\ \bibinfo
  {pages} {004}},\ \Eprint {https://arxiv.org/abs/2311.17990} {arXiv:2311.17990
  [hep-th]} \BibitemShut {NoStop}%
\bibitem [{\citenamefont {Bzowski}\ \emph {et~al.}(2023)\citenamefont
  {Bzowski}, \citenamefont {McFadden},\ and\ \citenamefont
  {Skenderis}}]{Bzowski:2023nef}%
  \BibitemOpen
  \bibfield  {author} {\bibinfo {author} {\bibfnamefont {A.}~\bibnamefont
  {Bzowski}}, \bibinfo {author} {\bibfnamefont {P.}~\bibnamefont {McFadden}},\
  and\ \bibinfo {author} {\bibfnamefont {K.}~\bibnamefont {Skenderis}},\
  }\bibfield  {title} {\bibinfo {title} {{Renormalisation of IR divergences and
  holography in de Sitter}},\ }\href@noop {} {\  (\bibinfo {year} {2023})},\
  \Eprint {https://arxiv.org/abs/2312.17316} {arXiv:2312.17316 [hep-th]}
  \BibitemShut {NoStop}%
\bibitem [{\citenamefont {Gorbenko}\ and\ \citenamefont
  {Senatore}(2019)}]{Gorbenko:2019rza}%
  \BibitemOpen
  \bibfield  {author} {\bibinfo {author} {\bibfnamefont {V.}~\bibnamefont
  {Gorbenko}}\ and\ \bibinfo {author} {\bibfnamefont {L.}~\bibnamefont
  {Senatore}},\ }\bibfield  {title} {\bibinfo {title} {{$\lambda \phi^4$ in
  dS}},\ }\href@noop {} {\  (\bibinfo {year} {2019})},\ \Eprint
  {https://arxiv.org/abs/1911.00022} {arXiv:1911.00022 [hep-th]} \BibitemShut
  {NoStop}%
\bibitem [{\citenamefont {Maldacena}\ and\ \citenamefont
  {Pimentel}(2011)}]{Maldacena:2011nz}%
  \BibitemOpen
  \bibfield  {author} {\bibinfo {author} {\bibfnamefont {J.~M.}\ \bibnamefont
  {Maldacena}}\ and\ \bibinfo {author} {\bibfnamefont {G.~L.}\ \bibnamefont
  {Pimentel}},\ }\bibfield  {title} {\bibinfo {title} {{On graviton
  non-Gaussianities during inflation}},\ }\href
  {https://doi.org/10.1007/JHEP09(2011)045} {\bibfield  {journal} {\bibinfo
  {journal} {JHEP}\ }\textbf {\bibinfo {volume} {09}},\ \bibinfo {pages}
  {045}},\ \Eprint {https://arxiv.org/abs/1104.2846} {arXiv:1104.2846 [hep-th]}
  \BibitemShut {NoStop}%
\bibitem [{\citenamefont {Raju}(2012)}]{Raju:2012zr}%
  \BibitemOpen
  \bibfield  {author} {\bibinfo {author} {\bibfnamefont {S.}~\bibnamefont
  {Raju}},\ }\bibfield  {title} {\bibinfo {title} {{New Recursion Relations and
  a Flat Space Limit for AdS/CFT Correlators}},\ }\href
  {https://doi.org/10.1103/PhysRevD.85.126009} {\bibfield  {journal} {\bibinfo
  {journal} {Phys. Rev. D}\ }\textbf {\bibinfo {volume} {85}},\ \bibinfo
  {pages} {126009} (\bibinfo {year} {2012})},\ \Eprint
  {https://arxiv.org/abs/1201.6449} {arXiv:1201.6449 [hep-th]} \BibitemShut
  {NoStop}%
\bibitem [{\citenamefont {Pajer}\ \emph {et~al.}(2020)\citenamefont {Pajer},
  \citenamefont {Stefanyszyn},\ and\ \citenamefont {Supe\l{}}}]{Pajer:2020wnj}%
  \BibitemOpen
  \bibfield  {author} {\bibinfo {author} {\bibfnamefont {E.}~\bibnamefont
  {Pajer}}, \bibinfo {author} {\bibfnamefont {D.}~\bibnamefont {Stefanyszyn}},\
  and\ \bibinfo {author} {\bibfnamefont {J.}~\bibnamefont {Supe\l{}}},\
  }\bibfield  {title} {\bibinfo {title} {{The Boostless Bootstrap: Amplitudes
  without Lorentz boosts}},\ }\href {https://doi.org/10.1007/JHEP12(2020)198}
  {\bibfield  {journal} {\bibinfo  {journal} {JHEP}\ }\textbf {\bibinfo
  {volume} {12}},\ \bibinfo {pages} {198}},\ \bibinfo {note} {[Erratum: JHEP
  04, 023 (2022)]},\ \Eprint {https://arxiv.org/abs/2007.00027}
  {arXiv:2007.00027 [hep-th]} \BibitemShut {NoStop}%
\bibitem [{\citenamefont {Benincasa}\ and\ \citenamefont
  {Vaz\~ao}(2024)}]{Benincasa:2024lxe}%
  \BibitemOpen
  \bibfield  {author} {\bibinfo {author} {\bibfnamefont {P.}~\bibnamefont
  {Benincasa}}\ and\ \bibinfo {author} {\bibfnamefont {F.}~\bibnamefont
  {Vaz\~ao}},\ }\bibfield  {title} {\bibinfo {title} {{The Asymptotic Structure
  of Cosmological Integrals}},\ }\href@noop {} {\  (\bibinfo {year} {2024})},\
  \Eprint {https://arxiv.org/abs/2402.06558} {arXiv:2402.06558 [hep-th]}
  \BibitemShut {NoStop}%
\bibitem [{\citenamefont {Benincasa}\ and\ \citenamefont
  {Dian}(2024)}]{Benincasa:2024leu}%
  \BibitemOpen
  \bibfield  {author} {\bibinfo {author} {\bibfnamefont {P.}~\bibnamefont
  {Benincasa}}\ and\ \bibinfo {author} {\bibfnamefont {G.}~\bibnamefont
  {Dian}},\ }\bibfield  {title} {\bibinfo {title} {{The Geometry of
  Cosmological Correlators}},\ }\href@noop {} {\  (\bibinfo {year} {2024})},\
  \Eprint {https://arxiv.org/abs/2401.05207} {arXiv:2401.05207 [hep-th]}
  \BibitemShut {NoStop}%
\bibitem [{\citenamefont {Arkani-Hamed}\ \emph {et~al.}(2023)\citenamefont
  {Arkani-Hamed}, \citenamefont {Baumann}, \citenamefont {Hillman},
  \citenamefont {Joyce}, \citenamefont {Lee},\ and\ \citenamefont
  {Pimentel}}]{Arkani-Hamed:2023kig}%
  \BibitemOpen
  \bibfield  {author} {\bibinfo {author} {\bibfnamefont {N.}~\bibnamefont
  {Arkani-Hamed}}, \bibinfo {author} {\bibfnamefont {D.}~\bibnamefont
  {Baumann}}, \bibinfo {author} {\bibfnamefont {A.}~\bibnamefont {Hillman}},
  \bibinfo {author} {\bibfnamefont {A.}~\bibnamefont {Joyce}}, \bibinfo
  {author} {\bibfnamefont {H.}~\bibnamefont {Lee}},\ and\ \bibinfo {author}
  {\bibfnamefont {G.~L.}\ \bibnamefont {Pimentel}},\ }\bibfield  {title}
  {\bibinfo {title} {{Differential Equations for Cosmological Correlators}},\
  }\href@noop {} {\  (\bibinfo {year} {2023})},\ \Eprint
  {https://arxiv.org/abs/2312.05303} {arXiv:2312.05303 [hep-th]} \BibitemShut
  {NoStop}%
\bibitem [{\citenamefont {Stefanyszyn}\ \emph {et~al.}(2024)\citenamefont
  {Stefanyszyn}, \citenamefont {Tong},\ and\ \citenamefont
  {Zhu}}]{Stefanyszyn:2023qov}%
  \BibitemOpen
  \bibfield  {author} {\bibinfo {author} {\bibfnamefont {D.}~\bibnamefont
  {Stefanyszyn}}, \bibinfo {author} {\bibfnamefont {X.}~\bibnamefont {Tong}},\
  and\ \bibinfo {author} {\bibfnamefont {Y.}~\bibnamefont {Zhu}},\ }\bibfield
  {title} {\bibinfo {title} {{Cosmological correlators through the looking
  glass: reality, parity, and factorisation}},\ }\href
  {https://doi.org/10.1007/JHEP05(2024)196} {\bibfield  {journal} {\bibinfo
  {journal} {JHEP}\ }\textbf {\bibinfo {volume} {05}},\ \bibinfo {pages}
  {196}},\ \Eprint {https://arxiv.org/abs/2309.07769} {arXiv:2309.07769
  [hep-th]} \BibitemShut {NoStop}%
\bibitem [{\citenamefont {Tong}\ \emph {et~al.}(2022)\citenamefont {Tong},
  \citenamefont {Wang},\ and\ \citenamefont {Zhu}}]{Tong:2021wai}%
  \BibitemOpen
  \bibfield  {author} {\bibinfo {author} {\bibfnamefont {X.}~\bibnamefont
  {Tong}}, \bibinfo {author} {\bibfnamefont {Y.}~\bibnamefont {Wang}},\ and\
  \bibinfo {author} {\bibfnamefont {Y.}~\bibnamefont {Zhu}},\ }\bibfield
  {title} {\bibinfo {title} {{Cutting rule for cosmological collider signals: a
  bulk evolution perspective}},\ }\href
  {https://doi.org/10.1007/JHEP03(2022)181} {\bibfield  {journal} {\bibinfo
  {journal} {JHEP}\ }\textbf {\bibinfo {volume} {03}},\ \bibinfo {pages}
  {181}},\ \Eprint {https://arxiv.org/abs/2112.03448} {arXiv:2112.03448
  [hep-th]} \BibitemShut {NoStop}%
\bibitem [{\citenamefont {Ema}\ and\ \citenamefont
  {Mukaida}(2024)}]{Ema:2024hkj}%
  \BibitemOpen
  \bibfield  {author} {\bibinfo {author} {\bibfnamefont {Y.}~\bibnamefont
  {Ema}}\ and\ \bibinfo {author} {\bibfnamefont {K.}~\bibnamefont {Mukaida}},\
  }\bibfield  {title} {\bibinfo {title} {{Cutting rule for in-in correlators
  and cosmological collider}},\ }\href@noop {} {\  (\bibinfo {year} {2024})},\
  \Eprint {https://arxiv.org/abs/2409.07521} {arXiv:2409.07521 [hep-th]}
  \BibitemShut {NoStop}%
\bibitem [{\citenamefont {Qin}\ and\ \citenamefont
  {Xianyu}(2023)}]{Qin:2023bjk}%
  \BibitemOpen
  \bibfield  {author} {\bibinfo {author} {\bibfnamefont {Z.}~\bibnamefont
  {Qin}}\ and\ \bibinfo {author} {\bibfnamefont {Z.-Z.}\ \bibnamefont
  {Xianyu}},\ }\bibfield  {title} {\bibinfo {title} {{Inflation correlators at
  the one-loop order: nonanalyticity, factorization, cutting rule, and OPE}},\
  }\href {https://doi.org/10.1007/JHEP09(2023)116} {\bibfield  {journal}
  {\bibinfo  {journal} {JHEP}\ }\textbf {\bibinfo {volume} {09}},\ \bibinfo
  {pages} {116}},\ \Eprint {https://arxiv.org/abs/2304.13295} {arXiv:2304.13295
  [hep-th]} \BibitemShut {NoStop}%
\bibitem [{\citenamefont {Qin}\ and\ \citenamefont
  {Xianyu}(2024)}]{Qin:2023nhv}%
  \BibitemOpen
  \bibfield  {author} {\bibinfo {author} {\bibfnamefont {Z.}~\bibnamefont
  {Qin}}\ and\ \bibinfo {author} {\bibfnamefont {Z.-Z.}\ \bibnamefont
  {Xianyu}},\ }\bibfield  {title} {\bibinfo {title} {{Nonanalyticity and
  on-shell factorization of inflation correlators at all loop orders}},\ }\href
  {https://doi.org/10.1007/JHEP01(2024)168} {\bibfield  {journal} {\bibinfo
  {journal} {JHEP}\ }\textbf {\bibinfo {volume} {01}},\ \bibinfo {pages}
  {168}},\ \Eprint {https://arxiv.org/abs/2308.14802} {arXiv:2308.14802
  [hep-th]} \BibitemShut {NoStop}%
\bibitem [{\citenamefont {DeWitt}(1967)}]{DeWitt:1967ub}%
  \BibitemOpen
  \bibfield  {author} {\bibinfo {author} {\bibfnamefont {B.~S.}\ \bibnamefont
  {DeWitt}},\ }\bibfield  {title} {\bibinfo {title} {{Quantum Theory of
  Gravity. 2. The Manifestly Covariant Theory}},\ }\href
  {https://doi.org/10.1103/PhysRev.162.1195} {\bibfield  {journal} {\bibinfo
  {journal} {Phys. Rev.}\ }\textbf {\bibinfo {volume} {162}},\ \bibinfo {pages}
  {1195} (\bibinfo {year} {1967})}\BibitemShut {NoStop}%
\bibitem [{\citenamefont {Lee}\ \emph {et~al.}(2023)\citenamefont {Lee},
  \citenamefont {McCulloch},\ and\ \citenamefont {Pajer}}]{Lee:2023jby}%
  \BibitemOpen
  \bibfield  {author} {\bibinfo {author} {\bibfnamefont {M.~H.~G.}\
  \bibnamefont {Lee}}, \bibinfo {author} {\bibfnamefont {C.}~\bibnamefont
  {McCulloch}},\ and\ \bibinfo {author} {\bibfnamefont {E.}~\bibnamefont
  {Pajer}},\ }\bibfield  {title} {\bibinfo {title} {{Leading loops in
  cosmological correlators}},\ }\href {https://doi.org/10.1007/JHEP11(2023)038}
  {\bibfield  {journal} {\bibinfo  {journal} {JHEP}\ }\textbf {\bibinfo
  {volume} {11}},\ \bibinfo {pages} {038}},\ \Eprint
  {https://arxiv.org/abs/2305.11228} {arXiv:2305.11228 [hep-th]} \BibitemShut
  {NoStop}%
\bibitem [{\citenamefont {Sleight}\ and\ \citenamefont
  {Taronna}(2021{\natexlab{a}})}]{Sleight:2020obc}%
  \BibitemOpen
  \bibfield  {author} {\bibinfo {author} {\bibfnamefont {C.}~\bibnamefont
  {Sleight}}\ and\ \bibinfo {author} {\bibfnamefont {M.}~\bibnamefont
  {Taronna}},\ }\bibfield  {title} {\bibinfo {title} {{From AdS to dS
  exchanges: Spectral representation, Mellin amplitudes, and crossing}},\
  }\href {https://doi.org/10.1103/PhysRevD.104.L081902} {\bibfield  {journal}
  {\bibinfo  {journal} {Phys. Rev. D}\ }\textbf {\bibinfo {volume} {104}},\
  \bibinfo {pages} {L081902} (\bibinfo {year} {2021}{\natexlab{a}})},\ \Eprint
  {https://arxiv.org/abs/2007.09993} {arXiv:2007.09993 [hep-th]} \BibitemShut
  {NoStop}%
\bibitem [{\citenamefont {Sleight}\ and\ \citenamefont
  {Taronna}(2021{\natexlab{b}})}]{Sleight:2021plv}%
  \BibitemOpen
  \bibfield  {author} {\bibinfo {author} {\bibfnamefont {C.}~\bibnamefont
  {Sleight}}\ and\ \bibinfo {author} {\bibfnamefont {M.}~\bibnamefont
  {Taronna}},\ }\bibfield  {title} {\bibinfo {title} {{From dS to AdS and
  back}},\ }\href {https://doi.org/10.1007/JHEP12(2021)074} {\bibfield
  {journal} {\bibinfo  {journal} {JHEP}\ }\textbf {\bibinfo {volume} {12}},\
  \bibinfo {pages} {074}},\ \Eprint {https://arxiv.org/abs/2109.02725}
  {arXiv:2109.02725 [hep-th]} \BibitemShut {NoStop}%
\bibitem [{\citenamefont {Goodhew}\ \emph {et~al.}(2024)\citenamefont
  {Goodhew}, \citenamefont {Thavanesan},\ and\ \citenamefont
  {Wall}}]{Goodhew:2024eup}%
  \BibitemOpen
  \bibfield  {author} {\bibinfo {author} {\bibfnamefont {H.}~\bibnamefont
  {Goodhew}}, \bibinfo {author} {\bibfnamefont {A.}~\bibnamefont
  {Thavanesan}},\ and\ \bibinfo {author} {\bibfnamefont {A.~C.}\ \bibnamefont
  {Wall}},\ }\bibfield  {title} {\bibinfo {title} {{The Cosmological CPT
  Theorem}},\ }\href@noop {} {\  (\bibinfo {year} {2024})},\ \Eprint
  {https://arxiv.org/abs/2408.17406} {arXiv:2408.17406 [hep-th]} \BibitemShut
  {NoStop}%
\bibitem [{\citenamefont {Liu}\ \emph {et~al.}(2020)\citenamefont {Liu},
  \citenamefont {Tong}, \citenamefont {Wang},\ and\ \citenamefont
  {Xianyu}}]{Liu:2019fag}%
  \BibitemOpen
  \bibfield  {author} {\bibinfo {author} {\bibfnamefont {T.}~\bibnamefont
  {Liu}}, \bibinfo {author} {\bibfnamefont {X.}~\bibnamefont {Tong}}, \bibinfo
  {author} {\bibfnamefont {Y.}~\bibnamefont {Wang}},\ and\ \bibinfo {author}
  {\bibfnamefont {Z.-Z.}\ \bibnamefont {Xianyu}},\ }\bibfield  {title}
  {\bibinfo {title} {{Probing P and CP Violations on the Cosmological
  Collider}},\ }\href {https://doi.org/10.1007/JHEP04(2020)189} {\bibfield
  {journal} {\bibinfo  {journal} {JHEP}\ }\textbf {\bibinfo {volume} {04}},\
  \bibinfo {pages} {189}},\ \Eprint {https://arxiv.org/abs/1909.01819}
  {arXiv:1909.01819 [hep-ph]} \BibitemShut {NoStop}%
\bibitem [{\citenamefont {Cabass}\ \emph {et~al.}(2023)\citenamefont {Cabass},
  \citenamefont {Jazayeri}, \citenamefont {Pajer},\ and\ \citenamefont
  {Stefanyszyn}}]{Cabass:2022rhr}%
  \BibitemOpen
  \bibfield  {author} {\bibinfo {author} {\bibfnamefont {G.}~\bibnamefont
  {Cabass}}, \bibinfo {author} {\bibfnamefont {S.}~\bibnamefont {Jazayeri}},
  \bibinfo {author} {\bibfnamefont {E.}~\bibnamefont {Pajer}},\ and\ \bibinfo
  {author} {\bibfnamefont {D.}~\bibnamefont {Stefanyszyn}},\ }\bibfield
  {title} {\bibinfo {title} {{Parity violation in the scalar trispectrum: no-go
  theorems and yes-go examples}},\ }\href
  {https://doi.org/10.1007/JHEP02(2023)021} {\bibfield  {journal} {\bibinfo
  {journal} {JHEP}\ }\textbf {\bibinfo {volume} {02}},\ \bibinfo {pages}
  {021}},\ \Eprint {https://arxiv.org/abs/2210.02907} {arXiv:2210.02907
  [hep-th]} \BibitemShut {NoStop}%
\bibitem [{\citenamefont {Shiraishi}\ \emph {et~al.}(2011)\citenamefont
  {Shiraishi}, \citenamefont {Nitta},\ and\ \citenamefont
  {Yokoyama}}]{Shiraishi:2011st}%
  \BibitemOpen
  \bibfield  {author} {\bibinfo {author} {\bibfnamefont {M.}~\bibnamefont
  {Shiraishi}}, \bibinfo {author} {\bibfnamefont {D.}~\bibnamefont {Nitta}},\
  and\ \bibinfo {author} {\bibfnamefont {S.}~\bibnamefont {Yokoyama}},\
  }\bibfield  {title} {\bibinfo {title} {{Parity Violation of Gravitons in the
  CMB Bispectrum}},\ }\href {https://doi.org/10.1143/PTP.126.937} {\bibfield
  {journal} {\bibinfo  {journal} {Prog. Theor. Phys.}\ }\textbf {\bibinfo
  {volume} {126}},\ \bibinfo {pages} {937} (\bibinfo {year} {2011})},\ \Eprint
  {https://arxiv.org/abs/1108.0175} {arXiv:1108.0175 [astro-ph.CO]}
  \BibitemShut {NoStop}%
\bibitem [{\citenamefont {Creminelli}\ \emph {et~al.}(2014)\citenamefont
  {Creminelli}, \citenamefont {Gleyzes}, \citenamefont {Nore\~na},\ and\
  \citenamefont {Vernizzi}}]{Creminelli:2014wna}%
  \BibitemOpen
  \bibfield  {author} {\bibinfo {author} {\bibfnamefont {P.}~\bibnamefont
  {Creminelli}}, \bibinfo {author} {\bibfnamefont {J.}~\bibnamefont {Gleyzes}},
  \bibinfo {author} {\bibfnamefont {J.}~\bibnamefont {Nore\~na}},\ and\
  \bibinfo {author} {\bibfnamefont {F.}~\bibnamefont {Vernizzi}},\ }\bibfield
  {title} {\bibinfo {title} {{Resilience of the standard predictions for
  primordial tensor modes}},\ }\href
  {https://doi.org/10.1103/PhysRevLett.113.231301} {\bibfield  {journal}
  {\bibinfo  {journal} {Phys. Rev. Lett.}\ }\textbf {\bibinfo {volume} {113}},\
  \bibinfo {pages} {231301} (\bibinfo {year} {2014})},\ \Eprint
  {https://arxiv.org/abs/1407.8439} {arXiv:1407.8439 [astro-ph.CO]}
  \BibitemShut {NoStop}%
\bibitem [{\citenamefont {Cabass}\ \emph {et~al.}(2022)\citenamefont {Cabass},
  \citenamefont {Pajer}, \citenamefont {Stefanyszyn},\ and\ \citenamefont
  {Supe\l{}}}]{Cabass:2021fnw}%
  \BibitemOpen
  \bibfield  {author} {\bibinfo {author} {\bibfnamefont {G.}~\bibnamefont
  {Cabass}}, \bibinfo {author} {\bibfnamefont {E.}~\bibnamefont {Pajer}},
  \bibinfo {author} {\bibfnamefont {D.}~\bibnamefont {Stefanyszyn}},\ and\
  \bibinfo {author} {\bibfnamefont {J.}~\bibnamefont {Supe\l{}}},\ }\bibfield
  {title} {\bibinfo {title} {{Bootstrapping large graviton
  non-Gaussianities}},\ }\href {https://doi.org/10.1007/JHEP05(2022)077}
  {\bibfield  {journal} {\bibinfo  {journal} {JHEP}\ }\textbf {\bibinfo
  {volume} {05}},\ \bibinfo {pages} {077}},\ \Eprint
  {https://arxiv.org/abs/2109.10189} {arXiv:2109.10189 [hep-th]} \BibitemShut
  {NoStop}%
\bibitem [{\citenamefont {Bordin}\ and\ \citenamefont
  {Cabass}(2020)}]{Bordin:2020eui}%
  \BibitemOpen
  \bibfield  {author} {\bibinfo {author} {\bibfnamefont {L.}~\bibnamefont
  {Bordin}}\ and\ \bibinfo {author} {\bibfnamefont {G.}~\bibnamefont
  {Cabass}},\ }\bibfield  {title} {\bibinfo {title} {{Graviton
  non-Gaussianities and Parity Violation in the EFT of Inflation}},\ }\href
  {https://doi.org/10.1088/1475-7516/2020/07/014} {\bibfield  {journal}
  {\bibinfo  {journal} {JCAP}\ }\textbf {\bibinfo {volume} {07}},\ \bibinfo
  {pages} {014}},\ \Eprint {https://arxiv.org/abs/2004.00619} {arXiv:2004.00619
  [astro-ph.CO]} \BibitemShut {NoStop}%
\bibitem [{\citenamefont {Bartolo}\ and\ \citenamefont
  {Orlando}(2017)}]{Bartolo:2017szm}%
  \BibitemOpen
  \bibfield  {author} {\bibinfo {author} {\bibfnamefont {N.}~\bibnamefont
  {Bartolo}}\ and\ \bibinfo {author} {\bibfnamefont {G.}~\bibnamefont
  {Orlando}},\ }\bibfield  {title} {\bibinfo {title} {{Parity breaking
  signatures from a Chern-Simons coupling during inflation: the case of
  non-Gaussian gravitational waves}},\ }\href
  {https://doi.org/10.1088/1475-7516/2017/07/034} {\bibfield  {journal}
  {\bibinfo  {journal} {JCAP}\ }\textbf {\bibinfo {volume} {07}},\ \bibinfo
  {pages} {034}},\ \Eprint {https://arxiv.org/abs/1706.04627} {arXiv:1706.04627
  [astro-ph.CO]} \BibitemShut {NoStop}%
\bibitem [{\citenamefont {Hou}\ \emph {et~al.}(2023)\citenamefont {Hou},
  \citenamefont {Slepian},\ and\ \citenamefont {Cahn}}]{Hou:2022wfj}%
  \BibitemOpen
  \bibfield  {author} {\bibinfo {author} {\bibfnamefont {J.}~\bibnamefont
  {Hou}}, \bibinfo {author} {\bibfnamefont {Z.}~\bibnamefont {Slepian}},\ and\
  \bibinfo {author} {\bibfnamefont {R.~N.}\ \bibnamefont {Cahn}},\ }\bibfield
  {title} {\bibinfo {title} {{Measurement of parity-odd modes in the
  large-scale 4-point correlation function of Sloan Digital Sky Survey Baryon
  Oscillation Spectroscopic Survey twelfth data release CMASS and LOWZ
  galaxies}},\ }\href {https://doi.org/10.1093/mnras/stad1062} {\bibfield
  {journal} {\bibinfo  {journal} {Mon. Not. Roy. Astron. Soc.}\ }\textbf
  {\bibinfo {volume} {522}},\ \bibinfo {pages} {5701} (\bibinfo {year}
  {2023})},\ \Eprint {https://arxiv.org/abs/2206.03625} {arXiv:2206.03625
  [astro-ph.CO]} \BibitemShut {NoStop}%
\bibitem [{\citenamefont {Philcox}(2022)}]{Philcox:2022hkh}%
  \BibitemOpen
  \bibfield  {author} {\bibinfo {author} {\bibfnamefont {O.~H.~E.}\
  \bibnamefont {Philcox}},\ }\bibfield  {title} {\bibinfo {title} {{Probing
  parity violation with the four-point correlation function of BOSS
  galaxies}},\ }\href {https://doi.org/10.1103/PhysRevD.106.063501} {\bibfield
  {journal} {\bibinfo  {journal} {Phys. Rev. D}\ }\textbf {\bibinfo {volume}
  {106}},\ \bibinfo {pages} {063501} (\bibinfo {year} {2022})},\ \Eprint
  {https://arxiv.org/abs/2206.04227} {arXiv:2206.04227 [astro-ph.CO]}
  \BibitemShut {NoStop}%
\bibitem [{\citenamefont {Philcox}\ and\ \citenamefont
  {Shiraishi}(2023)}]{Philcox:2023ypl}%
  \BibitemOpen
  \bibfield  {author} {\bibinfo {author} {\bibfnamefont {O.~H.~E.}\
  \bibnamefont {Philcox}}\ and\ \bibinfo {author} {\bibfnamefont
  {M.}~\bibnamefont {Shiraishi}},\ }\bibfield  {title} {\bibinfo {title}
  {{Testing Parity Symmetry with the Polarized Cosmic Microwave Background}},\
  }\href@noop {} {\  (\bibinfo {year} {2023})},\ \Eprint
  {https://arxiv.org/abs/2308.03831} {arXiv:2308.03831 [astro-ph.CO]}
  \BibitemShut {NoStop}%
\bibitem [{\citenamefont {Paul}\ \emph {et~al.}(2024)\citenamefont {Paul},
  \citenamefont {Clarkson},\ and\ \citenamefont {Maartens}}]{Paul:2024uim}%
  \BibitemOpen
  \bibfield  {author} {\bibinfo {author} {\bibfnamefont {P.}~\bibnamefont
  {Paul}}, \bibinfo {author} {\bibfnamefont {C.}~\bibnamefont {Clarkson}},\
  and\ \bibinfo {author} {\bibfnamefont {R.}~\bibnamefont {Maartens}},\
  }\bibfield  {title} {\bibinfo {title} {{Parity violation in the observed
  galaxy trispectrum}},\ }\href@noop {} {\  (\bibinfo {year} {2024})},\ \Eprint
  {https://arxiv.org/abs/2402.16478} {arXiv:2402.16478 [astro-ph.CO]}
  \BibitemShut {NoStop}%
\bibitem [{\citenamefont {Creque-Sarbinowski}\ \emph
  {et~al.}(2023)\citenamefont {Creque-Sarbinowski}, \citenamefont {Alexander},
  \citenamefont {Kamionkowski},\ and\ \citenamefont
  {Philcox}}]{Creque-Sarbinowski:2023wmb}%
  \BibitemOpen
  \bibfield  {author} {\bibinfo {author} {\bibfnamefont {C.}~\bibnamefont
  {Creque-Sarbinowski}}, \bibinfo {author} {\bibfnamefont {S.}~\bibnamefont
  {Alexander}}, \bibinfo {author} {\bibfnamefont {M.}~\bibnamefont
  {Kamionkowski}},\ and\ \bibinfo {author} {\bibfnamefont {O.}~\bibnamefont
  {Philcox}},\ }\bibfield  {title} {\bibinfo {title} {{Parity-violating
  trispectrum from Chern-Simons gravity}},\ }\href
  {https://doi.org/10.1088/1475-7516/2023/11/029} {\bibfield  {journal}
  {\bibinfo  {journal} {JCAP}\ }\textbf {\bibinfo {volume} {11}},\ \bibinfo
  {pages} {029}},\ \Eprint {https://arxiv.org/abs/2303.04815} {arXiv:2303.04815
  [astro-ph.CO]} \BibitemShut {NoStop}%
\bibitem [{\citenamefont {Maleknejad}(2016)}]{Maleknejad:2016qjz}%
  \BibitemOpen
  \bibfield  {author} {\bibinfo {author} {\bibfnamefont {A.}~\bibnamefont
  {Maleknejad}},\ }\bibfield  {title} {\bibinfo {title} {{Axion Inflation with
  an SU(2) Gauge Field: Detectable Chiral Gravity Waves}},\ }\href
  {https://doi.org/10.1007/JHEP07(2016)104} {\bibfield  {journal} {\bibinfo
  {journal} {JHEP}\ }\textbf {\bibinfo {volume} {07}},\ \bibinfo {pages}
  {104}},\ \Eprint {https://arxiv.org/abs/1604.03327} {arXiv:1604.03327
  [hep-ph]} \BibitemShut {NoStop}%
\bibitem [{\citenamefont {Maleknejad}\ and\ \citenamefont
  {Komatsu}(2019)}]{Maleknejad:2018nxz}%
  \BibitemOpen
  \bibfield  {author} {\bibinfo {author} {\bibfnamefont {A.}~\bibnamefont
  {Maleknejad}}\ and\ \bibinfo {author} {\bibfnamefont {E.}~\bibnamefont
  {Komatsu}},\ }\bibfield  {title} {\bibinfo {title} {{Production and
  Backreaction of Spin-2 Particles of $SU(2)$ Gauge Field during Inflation}},\
  }\href {https://doi.org/10.1007/JHEP05(2019)174} {\bibfield  {journal}
  {\bibinfo  {journal} {JHEP}\ }\textbf {\bibinfo {volume} {05}},\ \bibinfo
  {pages} {174}},\ \Eprint {https://arxiv.org/abs/1808.09076} {arXiv:1808.09076
  [hep-ph]} \BibitemShut {NoStop}%
\bibitem [{\citenamefont {Chen}\ \emph
  {et~al.}(2017{\natexlab{b}})\citenamefont {Chen}, \citenamefont {Wang},\ and\
  \citenamefont {Xianyu}}]{Chen:2017ryl}%
  \BibitemOpen
  \bibfield  {author} {\bibinfo {author} {\bibfnamefont {X.}~\bibnamefont
  {Chen}}, \bibinfo {author} {\bibfnamefont {Y.}~\bibnamefont {Wang}},\ and\
  \bibinfo {author} {\bibfnamefont {Z.-Z.}\ \bibnamefont {Xianyu}},\ }\bibfield
   {title} {\bibinfo {title} {{Schwinger-Keldysh Diagrammatics for Primordial
  Perturbations}},\ }\href {https://doi.org/10.1088/1475-7516/2017/12/006}
  {\bibfield  {journal} {\bibinfo  {journal} {JCAP}\ }\textbf {\bibinfo
  {volume} {12}},\ \bibinfo {pages} {006}},\ \Eprint
  {https://arxiv.org/abs/1703.10166} {arXiv:1703.10166 [hep-th]} \BibitemShut
  {NoStop}%
\bibitem [{\citenamefont {Benincasa}\ and\ \citenamefont
  {Cachazo}(2007)}]{Benincasa:2007xk}%
  \BibitemOpen
  \bibfield  {author} {\bibinfo {author} {\bibfnamefont {P.}~\bibnamefont
  {Benincasa}}\ and\ \bibinfo {author} {\bibfnamefont {F.}~\bibnamefont
  {Cachazo}},\ }\bibfield  {title} {\bibinfo {title} {{Consistency Conditions
  on the S-Matrix of Massless Particles}},\ }\href@noop {} {\  (\bibinfo {year}
  {2007})},\ \Eprint {https://arxiv.org/abs/0705.4305} {arXiv:0705.4305
  [hep-th]} \BibitemShut {NoStop}%
\bibitem [{\citenamefont {Cutkosky}(1960)}]{Cutkosky:1960sp}%
  \BibitemOpen
  \bibfield  {author} {\bibinfo {author} {\bibfnamefont {R.~E.}\ \bibnamefont
  {Cutkosky}},\ }\bibfield  {title} {\bibinfo {title} {{Singularities and
  discontinuities of Feynman amplitudes}},\ }\href
  {https://doi.org/10.1063/1.1703676} {\bibfield  {journal} {\bibinfo
  {journal} {J. Math. Phys.}\ }\textbf {\bibinfo {volume} {1}},\ \bibinfo
  {pages} {429} (\bibinfo {year} {1960})}\BibitemShut {NoStop}%
\end{thebibliography}%

\end{document}